\documentclass[review,numbers,sort&compress]{elsarticle}

\usepackage{hyperref}
\usepackage{amssymb,amsmath}
\usepackage{wrapfig}
\usepackage{caption}
\usepackage[labelformat=simple,textfont=normalfont,singlelinecheck=off,justification=raggedright]{subcaption}

\usepackage{tabularx}
\usepackage[capitalise]{cleveref}
\usepackage[margin=1in]{geometry}

\usepackage[usenames]{color} 

\usepackage{xcolor}

\newcommand{\MDGrevise}[1]{\textcolor{black}{#1}}

\newcommand{\RJHrevise}[1]{\textcolor{black}{#1}}
\usepackage[normalem]{ulem}

\newcommand{\Peclet}[1]{\text{P\'eclet}}

\journal{Journal of Non-Newtonian Fluid Mechanics}

\begin{document}

\begin{frontmatter}


\title{Flow instabilities in circular Couette flow of wormlike micelle solutions with a reentrant flow curve}

\author{Richard J. Hommel}

\author{Michael D. Graham\corref{mycorrespondingauthor}}
\cortext[mycorrespondingauthor]{Corresponding author}
\ead{mdgraham@wisc.edu}

\address{Department of Chemical and Biological Engineering, University of Wisconsin - Madison, Madison, Wisconsin 53706, USA}

\begin{abstract}
In this work, we numerically investigate \MDGrevise{flow instabilities of inertialess} circular Couette flow of dilute wormlike micelle solutions. Using the reformulated reactive rod model (RRM-R) [Hommel and Graham, JNNFM \textbf{295} (2021) 104606], which treats micelles as rigid Brownian rods undergoing reversible scission and fusion in flow, we study the development and behavior of both vorticity banding and finger-like instabilities. \RJHrevise{In particular, we focus on solutions that exhibit reentrant  constitutive curves, in which there exists some region where the shear stress, $\tau$, has a multivalued relation to shear rate, $\dot{\gamma}$. We find that the radial dependence of the shear stress in circular Couette flow allows for solutions in which parts of the domain lie in the region of the flow curve where $\partial \tau /\partial \dot{\gamma} > 0$, while others lie in the region where $\partial \tau /\partial \dot{\gamma} < 0$; this mixed behavior can lead to complex flow instabilities that manifest as finger-like structures of elongated and anisotropically-oriented micelles. In 3D simulations we find that the initial instability is 2D in origin, and 3D finger-like structures arise through the axial instability of 2D sheets.} Finally, we show that the RRM-R can capture vorticity banding in narrow-gap circular Couette flow and that vorticity bands are linearly stable to perturbations.
\end{abstract}


\end{frontmatter}


\section{Introduction}
\label{sec:Introduction}
Surfactants are amphiphilic molecules consisting of hydrophilic head groups bonded to long hydrophobic tails; when dissolved in water at some concentration above the critical micelle concentration (CMC), surfactants self-assemble into larger aggregate structures. The geometry of these structures is dictated by the size, shape, and chemistry of the surfactant molecules as well as the temperature and salinity of the solution \cite{Israelachvili2011,Oelschlaeger2010,Lerouge2009,Cates2006}. One class of these aggregate structures are wormlike (or rodlike) micelles, which can display a large range of varied structure and behavior depending on the concentration regime. In the dilute regime, wormlike micelles (WLMs) exist as nearly rigid rods with persistence lengths in the range of $\sim \mathcal{O}(10-100\mathrm{nm})$ and contour lengths on a similar scale \cite{Berret1998,Von1998,Zou2014,Imae1989,Ohlendorf1986}. In the concentrated regime, WLMs can grow far beyond their persistence lengths to form entangled networks and branched structures, transitioning the solution into a highly viscoelastic gel-like phase \cite{Helgeson2010}. In intermediate semi-dilute regimes, the behavior of WLMs depends on the applied flow or forcing; at rest and low forcing (e.g., low shear rates) these WLMs often form entangled structures,  while under the application of stronger flows these structures are dismantled and the WLMs show behavior that more closely resembles the dilute regime. 

Wormlike micelles can be found in a wide variety of commercial products, such as detergents, coatings, and emulsifiers, as well as industrial processes, such as in environmentally friendly carrier fluids for oil recovery operations \cite{Yang2002}. WLM solutions are also of great practical interest because they can provide significant levels of drag reduction in the transport of turbulent fluids. Notably, the addition of small amounts of wormlike micelle-forming surfactants to turbulent flows can produce up to an $80\%$ reduction in turbulent drag, which is comparable to the drag-reducing capabilities of widely used polymer solutions \cite{Zakin2017,Virk1975,Zakin1996}. Additionally, polymer molecules are shredded into short chain constituent segments by high-shear regions and must be continually replaced to achieve consistent drag reduction; wormlike micelles, however, are self-assembling and thus can overcome this shredding by reassembling following any mechanical degradation. Despite these benefits, the adoption of WLMs as drag-reducing agents has been limited, remaining mostly confined to areas of Japan for use in closed-loop heating and cooling districts \cite{Saeki2011,Krope2010}. Some of this limitation stems from chemical considerations regarding the amphiphilic nature of surfactants, specifically that changing the solvent can affect the aggregation and structuring of these molecules \cite{Shrestha2011,Tung2007}; other limitation stems from the fact that the behavior and dynamics of these fluids in complex flows, and specifically the development of instabilities in these flows, is not well understood especially compared to the flow of polymer solutions. In this work, we aim to expand understanding of \MDGrevise{flow} instabilities in dilute WLM solutions and elucidate the mechanisms underlying these instabilities.

\begin{figure}
    \centering
		\vspace{-3mm}
        \includegraphics[width=.5\linewidth]{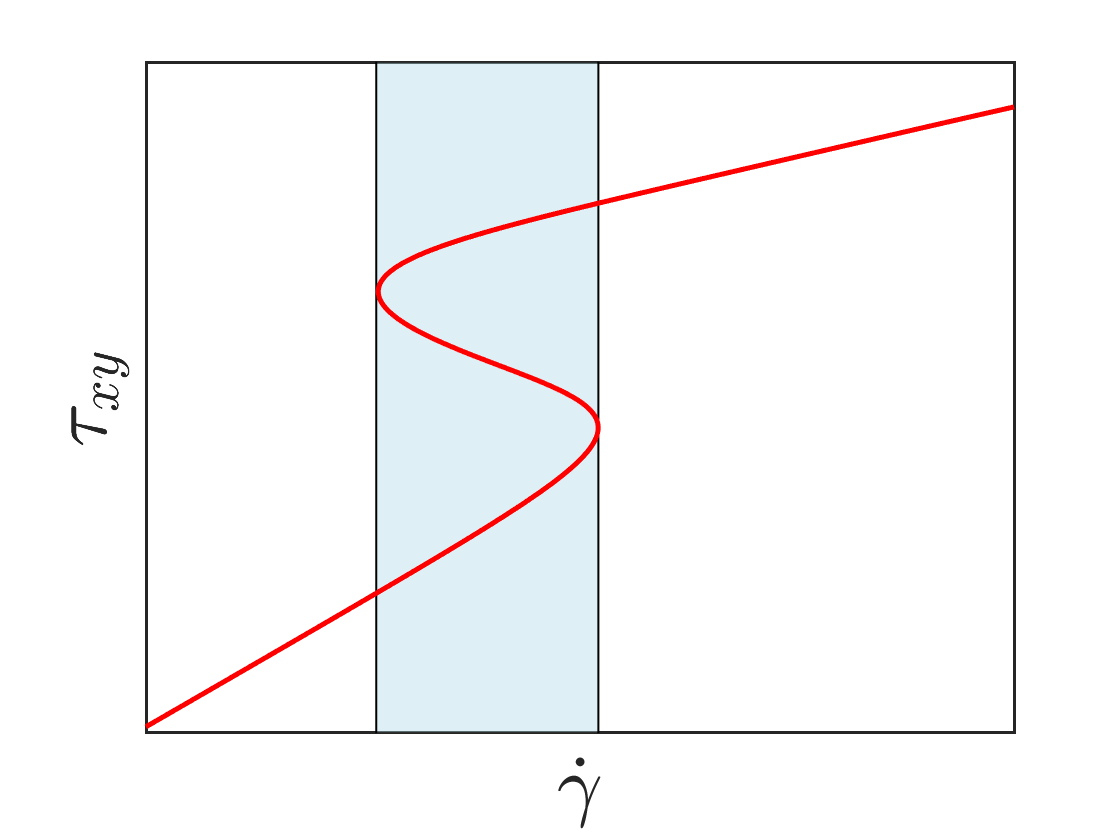}  
        \vspace{-3mm}
	    \caption{Example of a reentrant flow curve in which there is a region where the shear stress, $\tau_{xy}$, is a multivalued function of the shear rate, $\dot{\gamma}$. The light blue area shows the reentrant region of the flow curve.}
	    \label{fig:reentrantDiagram}
\end{figure}

Throughout this work we will focus exclusively on wormlike micelles in the dilute to semi-dilute regime. Wormlike micelle solutions in this regime are known to undergo both shear-thickening and shear-thinning, where thickening occurs at moderate shear rates followed by thinning at higher shear rates. This thickening and thinning behavior is related to the formation and subsequent breakdown of flow-induced structure (FIS). Specifically, at moderate shear rates WLMs are observed to align with the flow and undergo significant elongation, whereby the average length of micelles in solution can increase to several times the equilibrium length \cite{Lerouge2009,Keller1998}. This alignment and elongation leads to an increase in the viscosity of the solution. At higher shear rates, however, elongated micelles are broken down by the flow leading to a decrease in the viscosity. \cite{Perge2014,Wu2018,Mohammadigoushki2017,Fardin2012,Bhardwaj2007,Rojas2008,Berret1998}. These solutions are also well-known for displaying a reentrant, or multivalued, flow curve (see \cref{fig:reentrantDiagram}) whereby the shear stress becomes a multivalued function of shear rate over some typically small range \cite{Hommel2021,Boltenhagen1997,Herle2007}. 

The presence of a reentrant flow curve is a necessary condition for the development of a vorticity banding instability \cite{Herle2007}; this instability is a relatively uncommon phenomenon, but has been observed in charged colloidal suspensions, rodlike colloids, onion surfactants, biphasic polymer blends, and of course dilute WLM solutions \cite{Chen1992,Dhont2003a,Wilkins2006,Caserta2012,Herle2007,Mutze2014}. Vorticity banding is related to, but distinct from, the more well-studied shear banding instability where shear rate is a multivalued function of shear stress. Shear banding (also known as gradient banding) occurs at higher concentrations when micelles form entangled networks, and is characterized by the solution separating into a macroscopically ``banded'' flow along the gradient direction, where separate bands support equal shear stresses but different shear rates \cite{Dhont2008,Olmsted1999}. The separation of these two regions is often observable through differences in turbidity and birefringence \cite{Decruppe1995}. There has been extensive theoretical and experimental treatment of this instability (see \cite{Dhont2008,Olmsted2008} for comprehensive reviews). In contrast to gradient banding, vorticity banding requires that a single shear rate is able to support multiple shear stresses (i.e., a reentrant flow curve).  In circular Couette flow (CCF), this instability manifests as stacked ``bands'' along the vorticity axis, where adjacent bands support distinct shear stresses but equivalent shear rates. Again, similar to gradient banding, these bands can often be visualized by differences in turbidity and birefringence \cite{Dhont2008,Herle2007}.

\MDGrevise{The most basic stability analysis that is relevant to these situations is that of Yerushalmi et al. \cite{Yerushalmi1970}. They studied simple shear flow between parallel plates, with velocity field $v_x=\dot{\gamma}y$,  finding that, if $\partial \tau_{xy}/\partial \dot{\gamma} < 0$, the flow is linearly unstable to 1D ($y$-dependent) perturbations.}
\MDGrevise{This analysis does not distinguish between nonmonotonic and reentrant flow curves; it depends only on the local slope of the flow curve, not the global behavior. Below, when we describe the ``unstable" or ``stable" region of the flow curve, it is in the sense of this analysis. The actual nature of flow instabilities of a fluid that can display $\partial \tau_{xy}/\partial \dot{\gamma} < 0$ in simple shear depends on the global nature of both the constitutive behavior and flow geometry.} \MDGrevise{That said, the specific instability predicted by this analysis is 1D, and thus most directly related to gradient banding,} \RJHrevise{in which the shear rate is a multivalued function of shear stress,} though it has also been used to understand other instabilities as well \cite{Cromer2013,Divoux2016,Petrie1976,Olmsted2008}. In terms of vorticity \MDGrevise{banding} instabilities, most theoretical research has focused on  dense suspensions or through the construction of isotropic-nematic phase diagrams for rigid rods \cite{Olmsted1999,Dhont2008}. Olmsted and Lu \cite{Olmsted1999phase} modified the Doi model for rigid rod suspensions with inhomogeneous terms and found conditions in shear flow that could give rise to phase separation (i.e., banding). Fielding used the diffusive Johnson-Segalman model to show that gradient banded flow can be unstable to vorticity bands \cite{Fielding2007}. Chacko and coworkers \cite{Chacko2018} developed a simple continuum model for dense particle suspensions that displayed a constitutive curve with a region where $\partial \tau_{xy}/\partial \dot{\gamma} < 0$; they showed that the flow was linearly unstable to vorticity band formation, as expected, and further found that the vorticity banded flow was unsteady.

In addition to vorticity banding, dilute WLM solutions have demonstrated a group of finger-like instabilities in circular Couette flow that appear as unstructured streaks and branches spanning the flow gap \cite{Boltenhagen1997,Boltenhagen1997_2}. Using a circular Couette device, Liu and Pine \cite{Liu1996} observed these finger-like structures in controlled-shear rate flows of $\sim \mathcal{O} (10^2-10^3) ~ \text{ppm}$ equilmolar CTAB/NaSal solutions; using small angle light scattering (SALS), they revealed reversible finger-like structures that, upon increasing shear rate, grew outwards from the inner cylinder into the flow gap. The emergence of these structures required a finite induction time, typically on the order of minutes, and resulted in an elevated shear stress corresponding to significant shear-thickening. The authors observed that the fingers repeatedly grew until they reached the outer cylinder before retracting towards the center of the cell and beginning to grow again. Hu and coworkers \cite{Hu1998} observed similar structures in a solution of equimolar 7.5 mM TTAA/NaSal that exhibited a reentrant flow curve. Similar to Liu and Pine, the authors observed that increasing either the shear rate or shear stress beyond some critical value induced shear-thickened structures that originated at the inner cylinder and grew outwards to fill the gap. In stress-controlled experiments, they observed that increasing the applied stress into the multivalued region of the flow curve resulted in a FIS front that steadily grew to some radial position along the gap before stopping and holding at a steady gap location; the point along the gap where the FIS front stopped depended on the applied stress. When increasing the shear stress beyond the multivalued region, the authors observed that the FIS encompassed the entire gap. In shear rate-controlled experiments, the authors observed transient behavior that was similar to the stress-controlled experiments, but found that the steady behavior was different; notably the system was either void of FIS or exhibited FIS that filled the entire gap depending on whether the applied shear rate was below or above the critical value, respectively. 

Similar structures have also been observed to develop from interfacial instabilities. Wilson and Khomami have extensively studied interfacial instabilities in polymer melts and found that destabilization of these interfaces is strongly related to a jump in the first normal stress difference across the interface \cite{Wilson1992,Wilson1993_1,Wilson1993_2}. Pertinent to the current investigation is the structure of the interfacial instability, in which waves were bent and elongated by large shearing stresses, and in the case of compatible polymers `pools' of material are pulled off the wave to yield thread-like structures; these threads closely resemble the `fingers' observed in dilute WLM solution instabilities. Further, velocity gradients arising from interface curvature cause the waves to be convected at different rates, resulting in curved or 3D waves. Additionally, Yamani and coworkers \cite{Yamani2023} observed similar thread-like structures in their investigation of the flow of a planar jet of dilute polymer solution into a water tank. In their study, viscoelastic threads appeared to be sheared off of the main jet column. \RJHrevise{In the case of dilute wormlike micelle solutions, interface-like regions can manifest between domains of short, \MDGrevise{isotropically oriented} micelles and elongated, \MDGrevise{highly oriented} micelles.}

To the best of our knowledge, there have been no computational studies looking at the development of either vorticity bands or finger-like structures in shear flows of dilute WLM solutions. The main reason for this lack is the limited number of models for studying and predicting the behavior of WLM solutions in complex flows, and in particular models that can predict a reentrant flow curve. One of the first wormlike micelle models is due to Cates and Turner, who proposed a population balance model that accounted for the different stress relaxation mechanisms associated with wormlike micelles, namely micelle scission and rotational diffusion \cite{Turner1992,CatesTurner1990}. Though Cates and Turner did not write down an explicit constitutive equation for their model, and moreover the incorporation of a continuous spectrum of micelle lengths is prohibitive for use in computational fluid dynamics (CFD) studies, their formulation has served as a foundation for the development of other widely used WLM models \cite{Vasquez2007,Dutta2018}. Bautista and coworkers \cite{Bautista1999} have developed the BMP model, which couples a fluidity equation for studying thixotropic systems \cite{Fredrickson1970} to the Oldroyd-B equation. This model, as well as its many extensions and generalizations \cite{Manero2007,Lopez2018}, has shown good agreement with dilute wormlike micelles in a variety of flows \cite{Lopez2022}, and has recently been used for studying viscoelastoplastic and gradient banding fluids \cite{Lopez2022}. Tamano and coworkers \cite{Tamano2020} have taken inspiration from the BMP model and coupled the fluidity equation to both the Giesekus and FENE-P models to form the f-Giesekus and f-FENE-P models, respectively. These models are well-suited for CFD studies because many CFD codes and frameworks have already been developed for the FENE-P and Giesekus models \cite{Shekar2020,Dubief2023}, however, these models are unable to predict reentrant flow curves and are therefore unlikely to predict vorticity banding or finger-like instabilities. 

In this work, we investigate instability formation in circular Couette flows using the reformulated reactive rod model (RRM-R) \cite{Hommel2021}. The RRM-R, and its predecessor the RRM \cite{Dutta2018}, model wormlike micelles as reactive Brownian rods undergoing reversible scission and fusion in flow. The model couples evolution equations for the ensemble average orientation of rods and micelle contribution to the solution stress to an evolution equation for the collective length of micelles, where micelle number density and length are constrained by conservation of surfactant molecules. The evolution equation governing micelle length in the RRM-R, which was inspired by work by Turner and Cates \cite{Turner1992}, accounts for two forms of micelle fusion: spontaneous and flow-induced, as well as two forms of micelle scission: spontaneous and tension-induced. More details on the modeling framework of the RRM-R are provided in \cref{sec:ModelDescription}. Using this framework, the RRM-R can capture both shear-thickening and -thinning, flow-induced structure formation, nonzero normal stress differences, and importantly a reentrant flow curve. Moreover, we have shown in our previous work \cite{Hommel2021} that the RRM-R can be fit to experimental measurements of dilute WLM solution rheology, and can successfully predict the behavior of these solutions in both pure shear and pure extensional flows under both steady and transient conditions. This success in predicting the rheology of WLM solutions, along with the tractability of the RRM-R, makes this model well-suited for studying instability formation in complex flows. 


\section{Governing equations}
\label{sec:ModelDescription}
The aim of the present study is to investigate instabilities of dilute wormlike micelle solutions in circular Couette flow. Specifically, we are interested in exploring regimes where the underlying constitutive curve is reentrant as this region of state space is currently poorly understood and can give rise to interesting instabilities. To carry out our analysis we use the RRM-R (reactive rod model - reformulated), which models dilute WLM solutions as suspensions of reactive Brownian rods undergoing reversible scission and growth in flow. The RRM-R has shown qualitative agreement with experimental data of dilute WLM solutions in simple shear and purely extensional flows under both steady state and transient conditions.

The complete derivation of the RRM-R is described in \cite{Hommel2021} but is summarized below. We note that this modeling framework takes inspiration from theoretical treatments by Cates and Turner \cite{CatesTurner1990}. In the RRM-R, dilute wormlike micelle solutions are treated as suspensions of rigid Brownian rods undergoing reversible scission and fusion. Rods can fuse end-to-end (reducing the energetic penalty associated with the micellar end caps), but only when they are highly aligned -- otherwise the energy penalty arising from forming a long but bent micelle is too large for fusion to take place \cite{Larson1999,CatesTurner1990}. The application of flow tends to align the rods. This alignment is balanced by rotational diffusivity of the rods acting to return the suspension to isotropy. Consequently, a positive feedback mechanism exists between rod growth and alignment owing to the smaller rotational diffusivity of longer rods. It is assumed that rod growth is countered by hydrodynamic stresses acting along the lengths of the rods, which induce breakage events into shorter rods. Moreover, rods can undergo both spontaneous scission and spontaneous fusion events.

\subsection{Brownian rods}
Starting with a suspension of (non-reactive) Brownian rods, consider a uniform collection of rods with length $L_0$, radius $b$, and number density $n_0$ suspended in a Newtonian solvent with viscosity $\eta_s$. The orientation of a single rod is described by the unit director vector $\boldsymbol{u}$. The solution is subjected to an arbitrary flow with local velocity $\boldsymbol{v}$ and transpose velocity gradient $\boldsymbol{K} = \boldsymbol{\nabla}\boldsymbol{v}^\top$. The orientation tensor $\boldsymbol{S}$ describes the average collective orientation of the suspension and is given by the second moment of $\boldsymbol{u}$
\begin{equation}
    \boldsymbol{S} = \left\langle\boldsymbol{uu}\right\rangle = \int\boldsymbol{uu}\psi\mathrm{d}\boldsymbol{u},
\end{equation}
where $\psi$ is the probability distribution function of $\boldsymbol{u}$. The time evolution of $\boldsymbol{S}$ in flow is
\begin{equation}
    \frac{\mathrm{D}\boldsymbol{S}}{\mathrm{D}t} = -6D_{r,0}\left(\boldsymbol{S} - \frac{1}{3}\boldsymbol{I}\right) + \boldsymbol{K}\cdot\boldsymbol{S}^\top + \boldsymbol{S}\cdot\boldsymbol{K}^\top - 2\boldsymbol{K}:\left\langle\boldsymbol{uuuu}\right\rangle,
    \label{eqn:Orientation1}
\end{equation}
where $D_{r,0}$ is the rotational diffusion coefficient of a rod, $\boldsymbol{I}$ is the unit tensor, and the double dot product is defined as $\boldsymbol{A}:\boldsymbol{B} = \text{Tr}(\boldsymbol{A}\cdot\boldsymbol{B}^\top)$ \cite{Doi1986}.

The total stress of the suspension is given by the sum of the solvent $\boldsymbol{\tau}^N$ and micelle $\boldsymbol{\tau}^m$ contributions
\begin{equation}
    \boldsymbol{\tau}^T = \boldsymbol{\tau}^N + \boldsymbol{\tau}^m,
\end{equation}
where
\begin{equation}
    \boldsymbol{\tau}^N = 2\eta_s\boldsymbol{D}
\end{equation}
is the Newtonian solvent contribution with rate of deformation tensor $\boldsymbol{D} = \frac{1}{2}(\boldsymbol{K} + \boldsymbol{K}^\top)$ and
\begin{equation}
    \boldsymbol{\tau}^m = 3n_0k_BT\left(\boldsymbol{S}-\frac{1}{3}\boldsymbol{I}\right) + \frac{n_0k_BT}{2D_{r,0}}\boldsymbol{K}:\left\langle\boldsymbol{uuuu}\right\rangle
    \label{eqn:StressP}
\end{equation}
is the additional stress due to the presence of rods. Here, $k_B$ is the Boltzmann constant and $T$ is the temperature. \Cref{eqn:Orientation1,eqn:StressP} notably contain the fourth moment $\langle \boldsymbol{uuuu} \rangle$, an evolution equation for which depends on the sixth moment of $\boldsymbol{u}$, which in turn depends on higher moments. To proceed analytically, it is then necessary to supply a closure approximation for the product $\boldsymbol{K}:\left\langle\boldsymbol{uuuu}\right\rangle$. While numerous approximations are possible (see, for example: \cite{Doi1986,Dhont2006,Forest2003}), the RRM-R uses an approximation from Dhont and Briels \cite{Dhont2003} that interpolates between exact expressions in the limits of isotropy (equilibrium) and complete alignment:
\begin{gather}
    \boldsymbol{K}:\left\langle\boldsymbol{uuuu}\right\rangle \approx \frac{1}{5}[\boldsymbol{S}\cdot\boldsymbol{D}+\boldsymbol{D}\cdot\boldsymbol{S}-\boldsymbol{S}\cdot\boldsymbol{S}\cdot\boldsymbol{D}-\boldsymbol{D}\cdot\boldsymbol{S}\cdot\boldsymbol{S}+ 2\boldsymbol{S}\cdot\boldsymbol{D}\cdot\boldsymbol{S}+3(\boldsymbol{S}:\boldsymbol{D})\boldsymbol{S}].
    \label{eqn:Closure}
\end{gather}

\subsection{RRM-R}
As discussed above, a key feature of the RRM and RRM-R is that they allow micelles, modeled as rigid rods, to undergo reversible scission and growth by allowing the collective length and number density of the suspension to be dynamic properties that evolve with time and flow. This variation is mathematically achieved in the RRM-R by changing the constant rod length $L_0$ to the dynamic length $L$. To make analytical progress and ensure the tractability of the model we assume the system can be characterized by a single, representative length, $L$. Now consider a suspension of rods at equilibrium with number density $n_0$ and equilibrium length $L_0$; the radius $b$ of the rods is taken to be constant. The evolution of length $L$ and number density $n$ are constrained at all times by the surfactant mass balance
 \begin{equation}
nL = n_0L_0. 	\label{eq:surfactantbalance}
 \end{equation}

The rotational diffusion constant for a rod of length $L_0$ and radius $b$ is given by \cite{Doi1986,Graham:2018ty}
\begin{equation}
    D_{r,0} = \frac{3k_BT}{\pi\eta_sL_0^3}\ln\left(\frac{L_0}{2b}\right).
\end{equation}
In the RRM-R, the constant rotational diffusion coefficient of the simple rigid rod model is replaced by the length-dependent coefficient
\begin{equation}
    D_r = \frac{D_{r,0}}{L^{*3}}\left(\frac{\ln L^* + m}{m}\right),
    \label{eqn:DiffusionCoeff}
\end{equation}
where $L^* = L/L_0$ is the dimensionless micelle length and $m = \ln[L_0/(2b)]$ is a constant related to the initial aspect ratio of the rods. Substituting \cref{eqn:DiffusionCoeff} into \cref{eqn:Orientation1,eqn:StressP}, we find
\begin{equation}
    \frac{\mathrm{D}\boldsymbol{S}}{\mathrm{D}t} = -6D_{r}\left(\boldsymbol{S} - \frac{1}{3}\boldsymbol{I}\right) + \boldsymbol{K}\cdot\boldsymbol{S}^\top + \boldsymbol{S}\cdot\boldsymbol{K}^\top - 2\boldsymbol{K}:\left\langle\boldsymbol{uuuu}\right\rangle
    \label{eqn:RRM_Orientation_OG}
\end{equation}
and
\begin{equation}
    \boldsymbol{\tau}^m = 3nk_BT\left(\boldsymbol{S}-\frac{1}{3}\boldsymbol{I}\right) + \frac{nk_BT}{2D_{r}}\boldsymbol{K}:\left\langle\boldsymbol{uuuu}\right\rangle.
    \label{eqn:RRM_Stress_OG}
\end{equation}

The orientation of rods in the suspension is tracked by introducing a  scalar orientational order parameter
\begin{equation}
    \widehat{S} = \sqrt{\frac{3}{2}\widehat{\boldsymbol{S}}:\widehat{\boldsymbol{S}}},
\end{equation}
where $\widehat{\boldsymbol{S}} = \boldsymbol{S} - \frac{1}{3}\boldsymbol{I}$ is the traceless part of $\boldsymbol{S}$. This order parameter varies between $\widehat{S} = 0$ for isotropic rods and $\widehat{S} = 1$ for perfectly aligned rods. Note that the description and equations above are valid for both the original RRM and the reformulation (RRM-R), the only variation between the two models is in the length evolution equation, discussed below. 

To allow for variability of rod length the RRM-R assumes a length evolution equation that balances growth and breakdown of micelles
\begin{equation}
    \frac{\mathrm{D}L}{\mathrm{D}t} = R_g + R_b,
    \label{eqn:LengthNew}
\end{equation}
where $R_g \geq 0$ is the rate of micelle growth and $R_b \leq 0$ is the rate of micelle breakdown. As discussed previously, the RRM-R assumes two forms of growth: spontaneous and alignment-induced, and two forms of breakage: spontaneous and tension-induced. Again, the complete derivation can be found in \cite{Hommel2021}. After a number of simplifications involving the surfactant mass balance \cref{eq:surfactantbalance}, relating spontaneous effects that must balance at equilibrium,  we have the overall length evolution equation
\begin{equation}
    \frac{\mathrm{D}L}{\mathrm{D}t} = k_{b0}\left(\frac{L_0^3}{L^2} - L\right) + k_{ga}(n_0L_0)^2\frac{\widehat{S}^2}{L^2} - k_{bt}\left[\exp\left(\frac{a}{L_0}\frac{\boldsymbol{S}:\boldsymbol{\tau}^m}{n_0k_BT}\right)-1\right].
    \label{eqn:LengthDimensional}
\end{equation}
This equation contains four parameters - $k_{b0}$, $k_{ga}$, $k_{bt}$, and $a$, which are related to the spontaneous breakage, alignment-induced growth, tension-induced breakage, and scission energy of micelles, respectively. 

The RRM-R constitutive equations are coupled to conservation of mass and momentum
\begin{gather}
    \boldsymbol{\nabla} \cdot \boldsymbol{v} = 0,
\end{gather}
\begin{gather}
    \rho\frac{\mathrm{D}\boldsymbol{v}}{\mathrm{D}t} = -\boldsymbol{\nabla}p +\eta_s\nabla^2\boldsymbol{v} + \boldsymbol{\nabla}\cdot \boldsymbol{\tau}^m,
    \label{eqn:momentum}
\end{gather}
where $\boldsymbol{v}$ is the velocity, $p$ is the pressure, $\rho$ is the density of the fluid, $\eta_s$ is the solvent viscosity which is assumed to be Newtonian, and $\boldsymbol{\tau}^m$ is the micelle contribution to the fluid stress.

\subsection{Circular Couette flow}
\begin{figure}
    \centering
		\vspace{-3mm}
        \includegraphics[width=.5\linewidth]{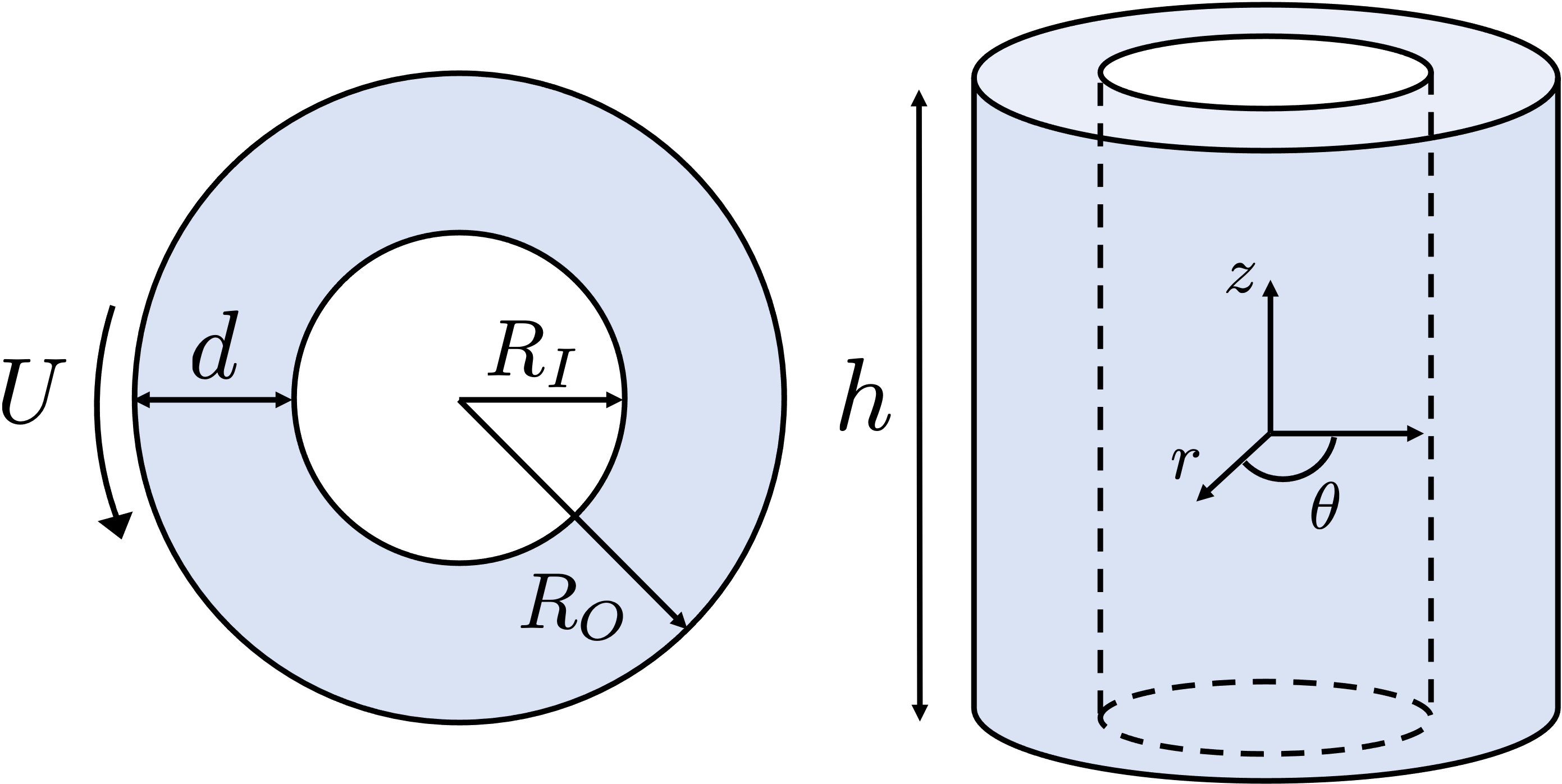}  
        \vspace{-3mm}
	    \caption{Schematic of circular Couette geometry. Left: top-down ($r\theta$) view of domain. Right: side view of domain.}
	    \label{fig:schematic}
\end{figure}

In this work we focus on the behavior of dilute wormlike micelle solutions in circular Couette flow, shown in \cref{fig:schematic}. The inner cylinder has radius $R_I$ and the outer cylinder has radius $R_O$; the gap width $d$ is the difference between the two, $d = R_O = R_I$; in all simulations we fix $d = 1$. The curvature ($\epsilon$) of the system is $\epsilon = d/R_I$. We take the inner cylinder to be stationary and the outer cylinder to rotate with some fixed angular velocity $\Omega_O$, which we can write as the linear azimuthal velocity $U=\Omega_0R_O$. The height of the cylinder is $h$. In 3D simulations we take all quantities to be periodic in $z$ at the cylinder ends. Using asterisks to denote dimensionless quantities, we render the governing equations dimensionless with the following relations: $\boldsymbol{x} = \boldsymbol{x}^*d$, $\boldsymbol{v} = \boldsymbol{v}^*U$, $t = t^*/\dot{\gamma}$, $p = p^*\eta_s \dot{\gamma}$, $\boldsymbol{\tau}^m = \boldsymbol{\tau}^{m*}G_0$, and $L = L^*L_0$ where $\dot{\gamma} = U/d$ is the characteristic shear rate and $G_0 = n_0k_BT = \eta_m D_{r,0}$ is the micelle shear modulus. Substituting these relations into the governing equations and dropping asterisks we are left with the dimensionless equations:
\begin{gather}
    \boldsymbol{\nabla} \cdot \boldsymbol{v} = 0,
\end{gather}
\begin{gather}
    \mathrm{Re}\frac{\mathrm{D}\boldsymbol{v}}{\mathrm{D}t} = -\boldsymbol{\nabla}p +\nabla^2\boldsymbol{v} + \left(\frac{1-\beta}{\beta}\right)\frac{1}{\mathrm{Pe}}\boldsymbol{\nabla}\cdot \boldsymbol{\tau}^m,
\end{gather}
\begin{equation}
    \frac{\mathrm{D}\boldsymbol{S}}{\mathrm{D}t} = -\frac{6}{\mathrm{Pe}L^3}\left(\boldsymbol{S} - \frac{1}{3}\boldsymbol{I}\right)\left(\frac{m + \ln L}{m}\right) + \boldsymbol{K}\cdot\boldsymbol{S}^\top + \boldsymbol{S}\cdot\boldsymbol{K}^\top - 2\boldsymbol{K}:\left\langle\boldsymbol{uuuu}\right\rangle,
    \label{eqn:S}
\end{equation}
\begin{equation}
    \boldsymbol{\tau}^m = \frac{3}{L}\left(\boldsymbol{S}-\frac{1}{3}\boldsymbol{I}\right) + \frac{\mathrm{Pe}}{2}\left(\frac{mL^2}{m + \ln L}\right)\boldsymbol{K}:\left\langle\boldsymbol{uuuu}\right\rangle,
    \label{eqn:tau}
\end{equation}
\begin{equation}
    \frac{\mathrm{D}L}{\mathrm{D}t} = \frac{1}{\mathrm{Pe}}\left[k_{b0}^*\left(\frac{1}{L^{2}} - L\right) + k_{ga}^*\frac{\widehat{S}^2}{L^{2}} - k_{bt}^*\left[\exp\left(a^*\boldsymbol{S}:\boldsymbol{\tau}^m\right)-1\right]\right].
    \label{eqn:L}
\end{equation}
We have introduced several dimensionless quantities. The Reynolds number is the ratio of inertial and viscous forces, defined as $\mathrm{Re} = \rho U d/\eta_s$. The rotational P\'eclet number is the ratio of the shear rate to the rotational diffusivity of the micelles at equilibrium, defined as $\mathrm{Pe} = \dot{\gamma}/D_{r0}$. In circular Couette flow the shear rate varies with radial position along the gap so that we have an \textit{applied} P\'eclet number defined with respect to the rotation rate of the outer cylinder, which we write as $\mathrm{Pe}$, and a \textit{local} P\'eclet number, written as $\mathrm{Pe}_l$, that is computed from the local shear rate in the gap. Finally, $\beta = \eta_s/(\eta_s+\eta_m)$ is the viscosity ratio. We also have four dimensionless groups in the length evolution equation -- $k_{b0}^*$, $k_{ga}^*$, $k_{bt}^*$, and $a^*$ -- all of which are defined identically to those in the previous work \cite{Hommel2021}. In order, $k_{b0}^*$ represents the ratio of relaxation due to spontaneous breakage to relaxation due to diffusion (i.e. realignment), $k_{ga}^*$ acts as a measure of the ratio of growth due to alignment to diffusion, $k_{bt}^*$ represents the ratio of relaxation due to tension-induced breakage to relaxation due to diffusion, and $a^*$ functions as a dimensionless length that must be overcome for tension-induced scission to occur. For the rest of this work we will drop the asterisks from these groups.

\subsection{Circular Couette flow: steady state}
To help analyze the behavior of the RRM-R in circular Couette flow (CCF), we will need to solve for the unidirectional steady state solutions of the system. We assume a purely azimuthal velocity profile that depends only on the radial coordinate, $\boldsymbol{v} = [0,v(r),0]^\top$. \RJHrevise{For this velocity profile the only non-zero component of the velocity gradient tensor is the $r\theta$ component, given by}
\begin{gather}
    \dot{\gamma}_{r\theta} = Dv - \frac{v}{r},
    \label{eqn:shearRate}
\end{gather}
where $D = \frac{\partial}{\partial r}$. For a unidirectional steady state velocity profile the equation for the $\theta$-component of the momentum equation \cref{eqn:momentum} is
\begin{gather}
    \left(r^2D^2 + rD - 1\right)v + \frac{1-\beta}{\beta}\frac{1}{\mathrm{Pe}}\left(2r + r^2D\right)\tau_{r\theta}^m = 0.
    \label{eqn:momentumComponents}
\end{gather}
We simplify the constitutive equations by writing the closure equation \cref{eqn:Closure} as
\begin{gather}
    \boldsymbol{C} = \boldsymbol{K}:\langle \boldsymbol{uuuu} \rangle.
    \label{eqn:ClosureSimplified}
\end{gather}
We then substitute the velocity profile, which is not yet known, and \cref{eqn:shearRate}  into \cref{eqn:ClosureSimplified} to obtain the components of the closure tensor
\begin{subequations}
    \begin{equation}
        C_{rr}=\frac{1}{5}S_{r\theta}\left(1-4S_{rr}-S_{\theta\theta}\right)\dot{\gamma}_{r\theta},
    \end{equation}
    \begin{equation}
        C_{r\theta}=-\frac{1}{10}\left(S_{rr}^2-(2S_{\theta\theta}+1)S_{rr} - 6S_{r\theta}^2 + S_{\theta\theta}^2 - S_{\theta\theta}\right)\dot{\gamma}_{r\theta},
    \end{equation}
    \begin{equation}
        C_{\theta\theta}=-\frac{1}{5}S_{r\theta}\left(-1+S_{rr}-4S_{\theta\theta}\right)\dot{\gamma}_{r\theta},
    \end{equation}
    \begin{equation}
        C_{zz}=\frac{3}{5}S_{r\theta}S_{zz}\dot{\gamma}_{r\theta}.
    \end{equation}
    \label[equation]{eqn:closureComponents}
\end{subequations}
We can then write the components of the orientation evolution equation
\begin{subequations}
    \begin{equation}
        \frac{\mathrm{D} S_{rr}}{\mathrm{D} t} = 0 = -\frac{6}{\mathrm{Pe}L^3}\left(\frac{\ln L^{*}+m}{m}\right)\left(S_{rr}-\frac{1}{3}\right)-2C_{rr},
    \end{equation}
    \begin{equation}
        \frac{\mathrm{D} S_{r\theta}}{\mathrm{D} t} = 0 = -\frac{6}{\mathrm{Pe}L^3}\left(\frac{\ln L^{*}+m}{m}\right)S_{r\theta} +S_{rr}\dot{\gamma}_{r\theta} -2C_{r\theta},
    \end{equation}
    \begin{equation}
       \frac{\mathrm{D} S_{\theta\theta}}{\mathrm{D} t} = 0 = -\frac{6}{\mathrm{Pe}L^3}\left(\frac{\ln L^{*}+m}{m}\right)\left(S_{\theta\theta}-\frac{1}{3}\right) + 2S_{r\theta}\dot{\gamma}_{r\theta} -2C_{\theta\theta},
    \end{equation}
    \begin{equation}
         \frac{\mathrm{D} S_{zz}}{\mathrm{D} t} = 0 = -\frac{6}{\mathrm{Pe}L^3}\left(\frac{\ln L^{*}+m}{m}\right)\left(S_{zz}-\frac{1}{3}\right) -2C_{zz}.
    \end{equation}
    \label[equation]{eqn:orientationComponents}
\end{subequations}
Likewise we have the components of the stress tensor
\begin{subequations}
    \begin{equation}
       \tau_{rr}^m=\frac{3}{L}\left(S_{rr}-\frac{1}{3}\right)+\frac{m \mathrm{Pe} L^2}{2\left(\ln L+m\right)}C_{rr},
    \end{equation}
    \begin{equation}
       \tau_{r\theta}^m=\frac{3}{L}S_{r\theta}+\frac{m \mathrm{Pe} L^2}{2\left(\ln L+m\right)}C_{r\theta},
    \end{equation}
    \begin{equation}
        \tau_{\theta\theta}^m=\frac{3}{L}\left(S_{\theta\theta}-\frac{1}{3}\right)+\frac{m \mathrm{Pe} L^2}{2\left(\ln L+m\right)}C_{\theta\theta},
    \end{equation}
    \begin{equation}
       \tau_{zz}^m=\frac{3}{L}\left(S_{zz}-\frac{1}{3}\right)+\frac{m \mathrm{Pe} L^2}{2\left(\ln L+m\right)}C_{zz},.
    \end{equation}
    \label[equation]{eqn:stressComponents}
\end{subequations}
In this flow the length evolution \cref{eqn:L} becomes
\begin{gather}
    \frac{\mathrm{D}L}{\mathrm{D}t} = 0 = k_{b0}\left(\frac{1}{L^{2}}-L\right) + k_{ga}\frac{\widehat{S}^2}{L^{2}} - k_{bt}\left[\exp\left(a\boldsymbol{S}:\boldsymbol{\tau}^m\right)-1\right],
    \label{eqn:lengthComponents}
\end{gather}
with
\begin{equation}
    \boldsymbol{S}:\boldsymbol{\tau}^m = S_{rr}\tau_{rr}^m + S_{\theta\theta}\tau_{\theta\theta}^m + S_{zz}\tau_{zz}^m + 2S_{r\theta}\tau_{r\theta}^m,
\end{equation}
and with scalar orientation parameter
\begin{equation}
    \widehat{S}=\left[\frac{3}{2}\left\{\left(S_{rr}-\frac{1}{3}\right)^{2}+\left(S_{\theta\theta}-\frac{1}{3}\right)^{2}+\left(S_{z z}-\frac{1}{3}\right)^{2}+2 S_{r \theta}^{2}\right\}\right]^{\frac{1}{2}}.
    \label{eqn:ShearShat}
\end{equation}


\section{Computational methods}
\label{sec:ComputationalMethods}

We solve the governing equations for mass, momentum, micelle orientation, and micelle length using the open-source CFD software OpenFOAM coupled with the viscoelastic solver RheoTool \cite{Jasak2007,Pimenta2017,rheoTool}. This framework uses the finite volume method to discretize equations. We have written an additional library for the RRM-R. Details of the numerical implementation of the code as well as validations are given elsewhere \cite{Favero2010,Pimenta2017}. In this study we perform both 2D and 3D numerical simulations of the RRM-R in circular Couette flow. For 3D simulations we take all quantities to be periodic in $z$ at the cylinder ends. In 2D simulations we confine the flow to the $r\theta$-plane and solve only the $rr$, $r\theta$, $\theta\theta$, and $zz$ components of the governing equations. On the cylinder walls we use no-slip and no-penetration boundary conditions for the velocity, zero normal gradient for the pressure, and linear-extrapolation conditions for the micelle length, orientation, and stress \cite{Pimenta2017}. Note that the stress boundary condition arises from numerical implementation, but in practice is fully determined by the length, orientation, and velocity of the fluid at the walls. To ensure numerical stability of our simulations we use the stress-velocity coupling method provided in RheoTool; since adding stabilization is known to alter transient dynamics, we incorporate a number of inner iterations to the main solver loop (typically 3-10 depending on the degree of stabilization added), which act to decrease the explicitness of the solver \cite{Pimenta2017}.

We generate the numerical grid for our problem using the \texttt{blockMesh} utility in OpenFOAM. To ensure the resolution of structures in our system and ensure mesh-independency of our solution, we tested four different 2D grid resolutions with densities: M0 = 40,000, M1 = 90,000, M2 = 250,000, and M3 = 640,000. In 3D, we use these same resolutions with 20-80 grid points in $z$. All results presented in this work use the M2 grid, unless otherwise stated. In general, we found that all resolutions showed quantitatively similar dynamics and results. Further, we tested the accuracy of our meshes by comparing time-dependent statistics and dynamics as well as steady state profiles to confirm that our solutions did not depend on the mesh density. 

To verify our viscoelastic library for the RRM-R and to look at a variety of both stable and unstable steady state solutions of our system, we also solve \cref{eqn:momentumComponents,eqn:closureComponents,eqn:orientationComponents,eqn:stressComponents,eqn:lengthComponents} numerically using a Chebyshev pseudospectral method on Gauss-Lobatto nodes. All results presented in this work use $N = 200$ nodes. We solve the discretized system of nonlinear equations using the \texttt{fsolve} solver in \textit{MATLAB}. We verified our codes by comparing steady state solutions obtained by the full DNS (OpenFOAM + RheoTool), the pseudospectral method (\textit{MATLAB}) assuming a purely azimuthal flow field, and the numerical solution to our equations in simple shear flow ($\epsilon = 0$). We tested multiple sets of RRM-R parameters at several different P\'eclet numbers and found agreement in all cases.


\section{Results and Discussion}
The organization of this section is as follows: In \Cref{sec:ss} we compute steady states for a reentrant flow curve and characterize the stability of these states at a variety of different curvatures. We then focus on 2D simulations with $\epsilon = 1$ in \Cref{sec:Finger_2D}, and show that unstable regions of the constitutive curve ($\partial \tau_{r\theta}/\partial \dot{\gamma} <0$) provoke finger-like instabilities. These finger-like structures are characterized by long branches of extended and anisotropically-oriented micelles. In \Cref{sec:Re_3D} we investigate the appearance of these finger-like structures in 3D and show that the underlying instability is 2D in nature. Finally, in \Cref{sec:Re_VB}, we briefly discuss and analyze the linear stability of vorticity banding in the reentrant system.

\subsection{Steady states in reentrant WLM solutions}
\label{sec:ss}
The RRM-R length evolution equation, \cref{eqn:L}, contains four dimensionless parameters that can be tuned to vary the behavior of the desired WLM system. We select values that yield a significantly reentrant constitutive curve so that we can adequately probe and characterize instabilities in this region. The values chosen for the constitutive model are: $m = 3$, $k_{b0} = 10^{-2}$, $k_{ga} = 1500$, $k_{bt} = 10$, and $a = 2.5$, with $\mathrm{Re} = 10^{-1}$ and $\beta = 0.4$. This value of $\beta$ may seem low for modeling dilute viscoelastic solutions, but was chosen to match typical values seen in experiments where the zero-shear viscosity of these solutions is between $4-10 ~ \text{mPa s}$ \cite{Liu1996,Ohlendorf1986}. Additionally, this value of $\mathrm{Re}$ is larger than what is typically used in experiments, but is small enough  such that inertial forces remain small without requiring an overly restrictive time step for stable time-integration.

\begin{figure}
    \centering
		\vspace{-3mm}
        \includegraphics[width=.8\linewidth]{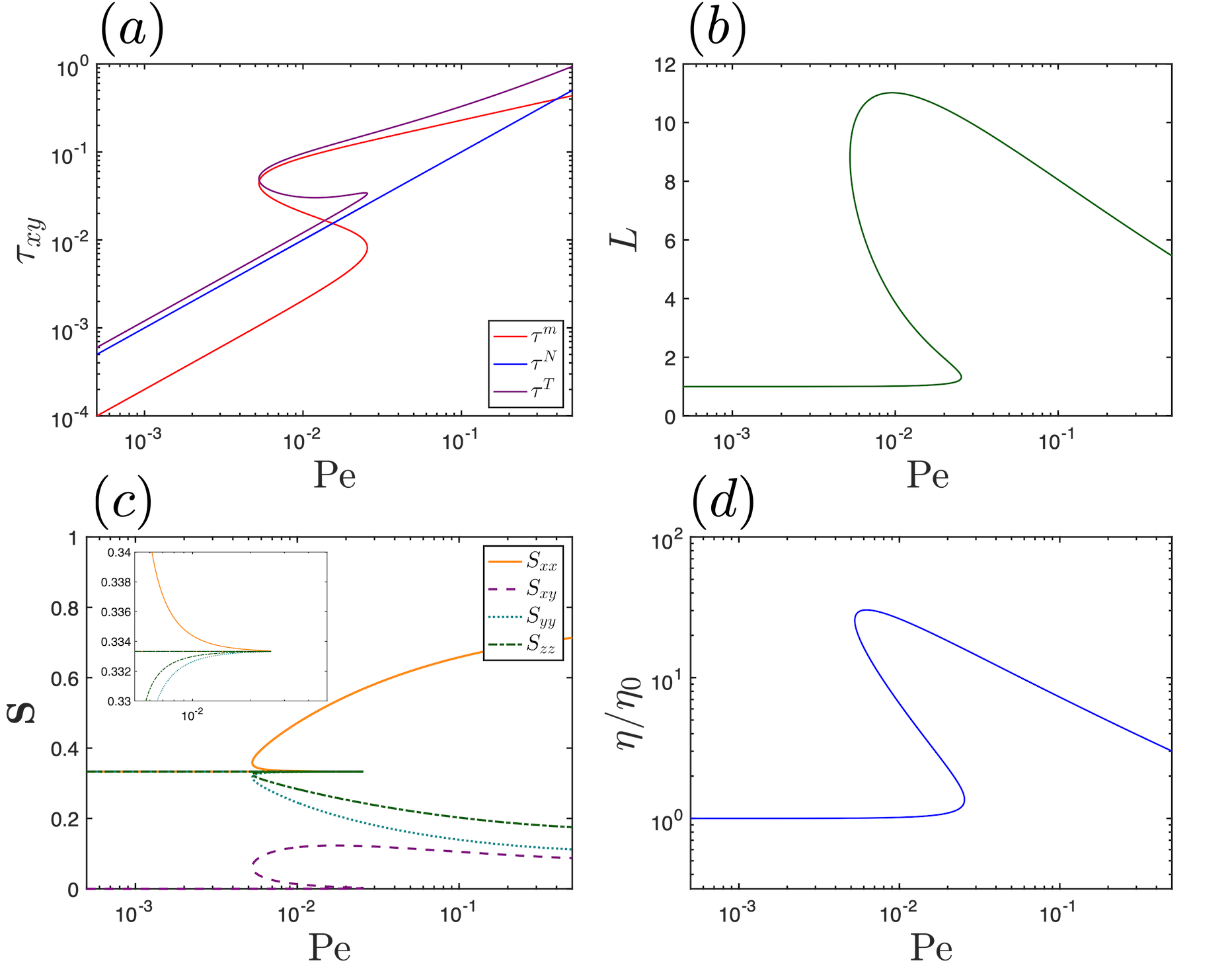}  
        \vspace{-3mm}
	    \caption{Constitutive curves for the RRM-R with parameters: $m = 3$, $k_{b0} = 10^{-2}$, $k_{ga} = 1500$, $k_{bt} = 10$, and $a = 2.5$. (a) Micelle (red), Newtonian (blue), and total shear stress (purple), (b) length of micelles, (c) components of the orientation tensor, and (d) normalized viscosity vs. applied P\'eclet number.}
	    \label{fig:constitutiveCurve1}
\end{figure}

In simple shear flow, $\boldsymbol{v} = (\mathrm{Pe}~y,0,0)^T$, these parameters yield the constitutive curves shown in \cref{fig:constitutiveCurve1}. These curves are highly reentrant over almost a decade of P\'eclet numbers, which will facilitate probing the behavior of this region, specifically sections where $\partial \tau_{xy}/\partial \dot{\gamma} <0$. The length of micelles for these parameters increases to a maximum of about 11 times the equilibrium length, and the solution shear-thickens by over an order of magnitude. The inset in \cref{fig:constitutiveCurve1}c shows a close-up of the orientation profiles near the turning point into the multivalued region. To clarify the analysis of these flows, we define three distinct solution branches on the shear stress constitutive curve (\cref{fig:constitutiveCurve1}a): the lower, middle, and upper branches with $\tau^T_{xy,low} < \tau^T_{xy,mid} < \tau^T_{xy,up}$. The lower branch extends from $0 < \mathrm{Pe} \lesssim 2.5 \times 10^{-2}$. The middle branch extends from $5 \times 10^{-3} \lesssim \mathrm{Pe} \lesssim 2.5 \times 10^{-2}$, where notably $\partial \tau^m_{xy}/\partial \dot{\gamma} < 0$ throughout this entire region. Finally, the upper branch extends from $5 \times 10^{-3} \lesssim \mathrm{Pe} < \infty$. 

In this work we are primarily interested in circular Couette flow. The notable difference between this flow and simple shear flow is that in CCF all quantities (e.g., stress, length, and orientation) vary across the gap due to the radial dependence of the shear rate. The precise radial dependence of these quantities depends on the underlying curvature of the system. We can understand the role of curvature through the steady state momentum balance, which simplifies to $\tau_{r\theta}^T(r) = \tau_{r\theta,I}^T(R_I/r)^2$, where $\tau_{r\theta,I}^T$ is the total shear stress at the inner cylinder. Rewriting this balance in terms of the curvature and outer cylinder stress ($\tau_{r\theta,O}^T$) we find $\tau_{r\theta,O}^T = \tau_{r\theta,I}^T/(1+\epsilon)^2$. Clearly, as the curvature decreases so does the difference between $\tau_{r\theta,O}^T$ and $\tau_{r\theta,I}^T$, leading to nearly constant properties across the gap. Conversely, as the curvature increases so does the difference between the stresses at the inner and outer cylinders. \RJHrevise{This radial dependence of the shear rate also complicates the characterization of steady state solutions for the reentrant flow curve; because the shear stress will vary throughout the gap, certain parts of the domain can lie in unstable regions of the flow curve ($\partial \tau_{r\theta}/\partial \dot{\gamma} < 0$) while the remainder lies in stable regions ($\partial \tau_{r\theta}/\partial \dot{\gamma} > 0$), leading to mixed local stability throughout the gap. Further, because the flow curve is multivalued over a range of P\'eclet numbers, more than one steady state can exist for a given applied $\mathrm{Pe}$.}

\begin{figure}
    \centering
		\vspace{-3mm}
        \includegraphics[width=0.95\linewidth]{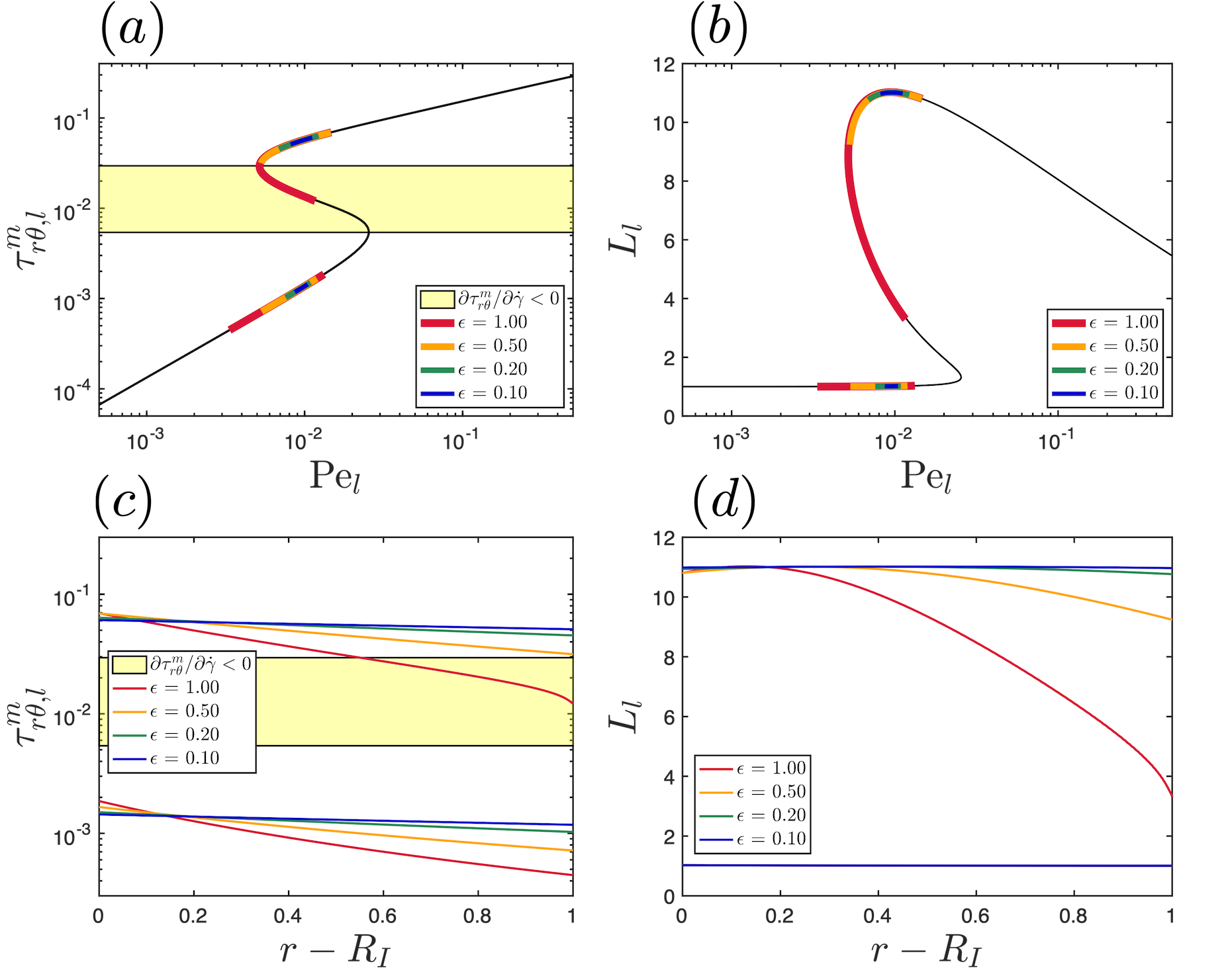}  
        \vspace{-3mm}
	    \caption{Upper and lower branch steady states for $\mathrm{Pe} = 0.01$ at varying curvatures. (a) Local micelle shear stress and (b) local micelle length projected onto the governing constitutive curves, where the local P\'eclet number, $\mathrm{Pe}_l$, is calculated from the velocity profile throughout the gap. (c) Local micelle shear stress and (d) local micelle length over the gap radius. Line colors are red: $\epsilon = 1.00$, orange: $\epsilon = 0.50$, green: $\epsilon = 0.20$, and blue: $\epsilon = 0.10$. The yellow rectangle indicates the unstable region  of the flow curve ($\partial \tau_{r\theta}^m/\partial \dot{\gamma} < 0$).}
	    \label{fig:SS_Pe01}
\end{figure}

To further emphasize the role of curvature and the fact that steady states can encompass both stable and unstable branches, we show the steady state solutions for $\mathrm{Pe} = 0.01$ at several curvatures in \cref{fig:SS_Pe01}. All steady states were found using the Chebyshev pseudospectral method described in \cref{sec:ComputationalMethods}. The top row shows the (a) local micelle shear stress and (b) local micelle length projected onto the governing constitutive curves where the local P\'eclet number, $\mathrm{Pe}_l$, is calculated from the velocity profile throughout the gap. The bottom row shows the (c) local micelle shear stress and (d) local micelle length over the gap radius. The line colors show different curvatures where red: $\epsilon = 1.00$, orange: $\epsilon = 0.50$, green: $\epsilon = 0.20$, and blue: $\epsilon = 0.10$. The yellow rectangle indicates the region that is locally unstable $(\partial \tau_{r\theta}^m/\partial \dot{\gamma} < 0)$. We only show the micelle stress, and not the total stress, in \cref{fig:SS_Pe01}a because the unstable region originates from the micelle contribution to the stress and anywhere where the micelle stress is multivalued so is the total stress. We can see from these plots that at $\mathrm{Pe} = 0.01$ two different steady states exist, one that is on the lower branch and one that is predominantly on the upper branch, but for larger curvatures the upper branch steady state extends into the middle branch. There is of course also a steady state that exists predominantly on the middle branch, however, this steady state \RJHrevise{is unstable and thus} challenging to observe both experimentally and computationally, so for the remainder of this work we will focus solely on the upper and lower branch steady states. For steady states on the lower branch, micelles do not exhibit any pronounced elongation and remain nearly at the equilibrium length. Also on the lower branch, the micelle shear stress at the inner cylinder ($r-R_I = 0$) increases with increasing curvature while at the outer cylinder the micelle shear stress decreases with increasing curvature. This observation will become important because it shows that the stability region on the lower branch, namely the range of P\'eclet numbers leading to a stable steady state, decreases with increasing curvature. The micelle shear stresses for all curvatures on this branch are equal around $r - R_I \approx 0.15$ where $\mathrm{Pe}_l \approx 0.01$.

On the upper branch we see that the micelle length varies significantly with curvature. In particular, for $\epsilon = 0.1$ and $\epsilon = 0.2$ the micelle length is nearly constant throughout the gap and micelles are nearly at the maximum degree of elongation prescribed by the constitutive curve. For the larger curvatures the length varies drastically throughout the gap; in the case of the largest curvature, $\epsilon = 1$, the length varies by almost an order of magnitude with maximum elongation occurring close to the inner cylinder. The large range of micelle lengths here results from the sharp change in length that occurs on the middle branch of the constitutive curve (\cref{fig:SS_Pe01}b). There is a similar trend in the micelle shear stress on the upper branch as there is on the lower branch, notably the micelle stress at the inner cylinder is largest for the highest curvature. Now, however, we also see that for $\epsilon = 1$ the micelle stress towards the outer cylinder is clearly falling into the unstable middle branch region where $\partial \tau_{r\theta}^m/\partial \dot{\gamma} < 0$, and therefore this steady state solution is unstable to inhomogeneous flow. The instability of this steady state emphasizes the role of curvature in dictating the stability of the system, in particular increasing curvature directly decreases the stability of the solution by increasing the span of shear stresses occupied throughout the gap. Moreover, for systems with low curvature (e.g., $\epsilon = 0.1$) the system can support two steady states that are both stable, which can then allow for the manifestation of vorticity bands. We will elaborate on vorticity banding towards the end of this work.

\begin{figure}
    \centering
		\vspace{-3mm}
        \includegraphics[width=0.95\linewidth]{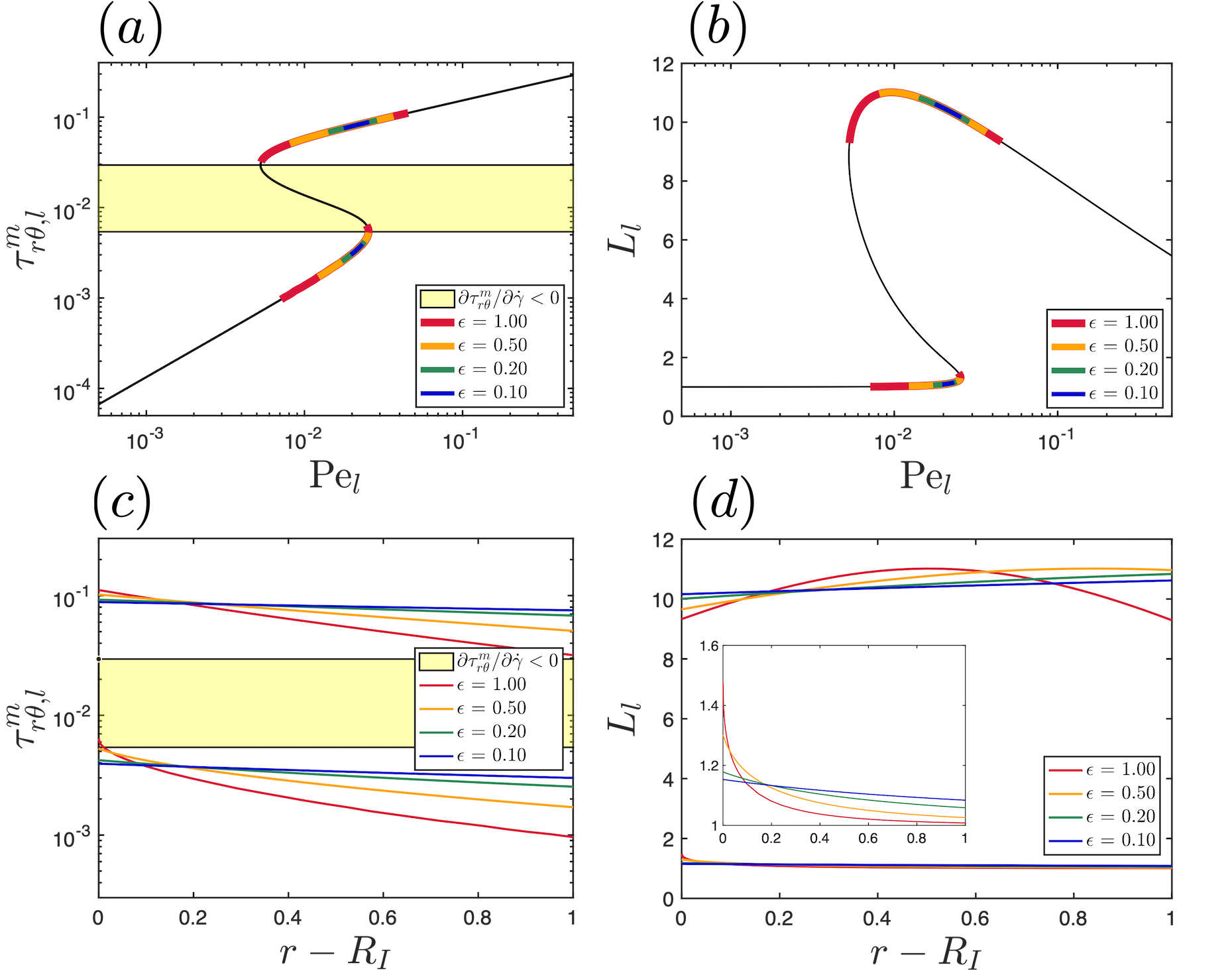}  
        \vspace{-3mm}
	    \caption{Upper and lower branch steady states for $\mathrm{Pe} = 0.0225$ at varying curvatures. (a) Local micelle shear stress and (b) local micelle length projected onto the governing constitutive curves, where the local P\'eclet number, $\mathrm{Pe}_l$, is calculated from the velocity profile throughout the gap. (c) Local micelle shear stress and (d) local micelle length over the gap radius. Line colors are red: $\epsilon = 1.00$, orange: $\epsilon = 0.50$, green: $\epsilon = 0.20$, and blue: $\epsilon = 0.10$. The yellow rectangle indicates the unstable region  of the flow curve ($\partial \tau_{r\theta}^m/\partial \dot{\gamma} < 0$).}
	    \label{fig:SS_Pe0225}
\end{figure}

\Cref{fig:SS_Pe0225} shows the same plots as \cref{fig:SS_Pe01} but now for $\mathrm{Pe} = 0.0225$. The inset in (d) shows a close-up of the micelle length. The observations for this Pe are very similar to the previous case of $\mathrm{Pe} = 0.01$. In (d) we see that for all curvatures on the upper branch micelles are nearly at the maximum degree of elongation throughout the entire gap. On the lower branch micelles are nearly completely at the equilibrium length, however, we see from the inset of (d) that for $\epsilon = 1$ the micelles grow very rapidly close to the inner cylinder where the flow is just beginning to enter the middle branch region. In (c) we see that the micelle shear stress at the inner cylinder increases with increasing curvature and this actually causes the micelle stress for $\epsilon = 1$ on the lower branch to enter into the middle region. Specifically, the transition from the lower branch into the unstable middle region occurs at the inner cylinder, whereas for Pe $=$ 0.01 it occurred at the outer cylinder. We then expect, and we will actually see later in this work, the instability at this Pe to originate close to the inner cylinder. 

\begin{figure}
    \centering
		\vspace{-3mm}
        \includegraphics[width=0.95\linewidth]{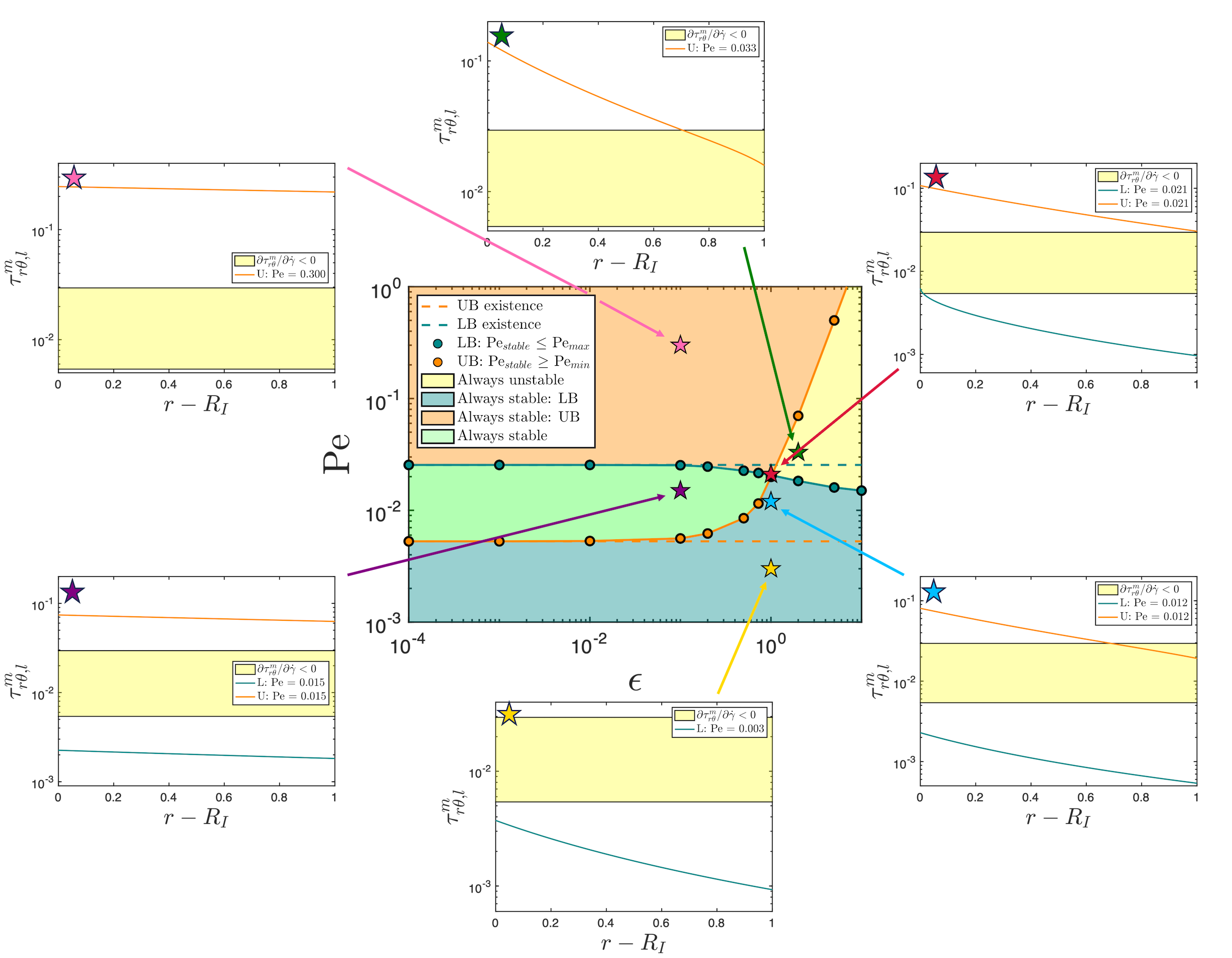}  
        \vspace{-3mm}
	    \caption{Shear rate-curvature state-space stability diagram. Dashed lines show the existence limits for the lower (cyan) and upper (orange) branches respectively. The cyan markers are the maximum Pe on the lower branch that will support a steady state; all Pe on the lower branch below this line are stable, shown by the cyan-shaded region. The orange markers are the minimum Pe on the upper branch that will support a steady state; all Pe on the upper branch above this line are stable, shown by the orange-shaded region. In the green-shaded region both the lower and upper branches exist and are stable. In the yellow-shaded region both the lower and upper branches are unstable or do not exist. The stars show the local shear stress profiles across the gap at the indicated Pe and $\epsilon$.}
	    \label{fig:bifurcation}
\end{figure}

We summarize the effects of curvature on the stability of steady states in the system with \cref{fig:bifurcation}. This figure shows the stability of the system for a given curvature and applied P\'eclet number. The dashed lines show the existence limits for the lower (cyan) and upper (orange) branches (i.e., Pe$_{max}$ on the lower branch and Pe$_{min}$ on the upper branch, beyond which the branches no longer exist). The cyan markers show the \textit{maximum} Pe value on the \textit{lower} branch that will support a stable steady state for a given curvature; all Pe below this value on the lower branch will be stable, shown by the cyan-shaded region. The orange markers show the \textit{minimum} Pe value on the \textit{upper} branch that will support a stable steady state for a given curvature; all Pe above this value on the upper branch will be stable, shown by the orange-shaded region. The green region shows the range of Pe and $\epsilon$ where both the lower and upper branches exist and are stable. The yellow region shows the range of Pe and $\epsilon$ where both branches are unstable or do not exist. The colored stars show the local micelle shear stress over the gap at the indicated Pe and $\epsilon$. The pink and green stars show locations where only the upper branch exists and is stable for the pink star and unstable for the green. The red star shows an unstable lower branch and a stable upper branch, though it is precariously close to the unstable region. The blue star shows a stable lower branch and unstable upper branch. The yellow star shows a location where only the lower branch exists and it is stable. Finally, the purple star shows a location where both lower and upper branches exist and are stable. We can see from \cref{fig:bifurcation} that the range of unstable Pe increases with increasing curvature, though the effects of curvature only start to become evident around $\epsilon \approx 10^{-1}$; for $\epsilon \lesssim 10^{-1}$, the Pe stability range is nearly independent of curvature. For $\epsilon \lesssim 1$, all Pe have some region where they are stable, whether this is on the lower branch or the upper branch.

\begin{figure}
    \centering
		\vspace{-3mm}
        \includegraphics[width=0.5\linewidth]{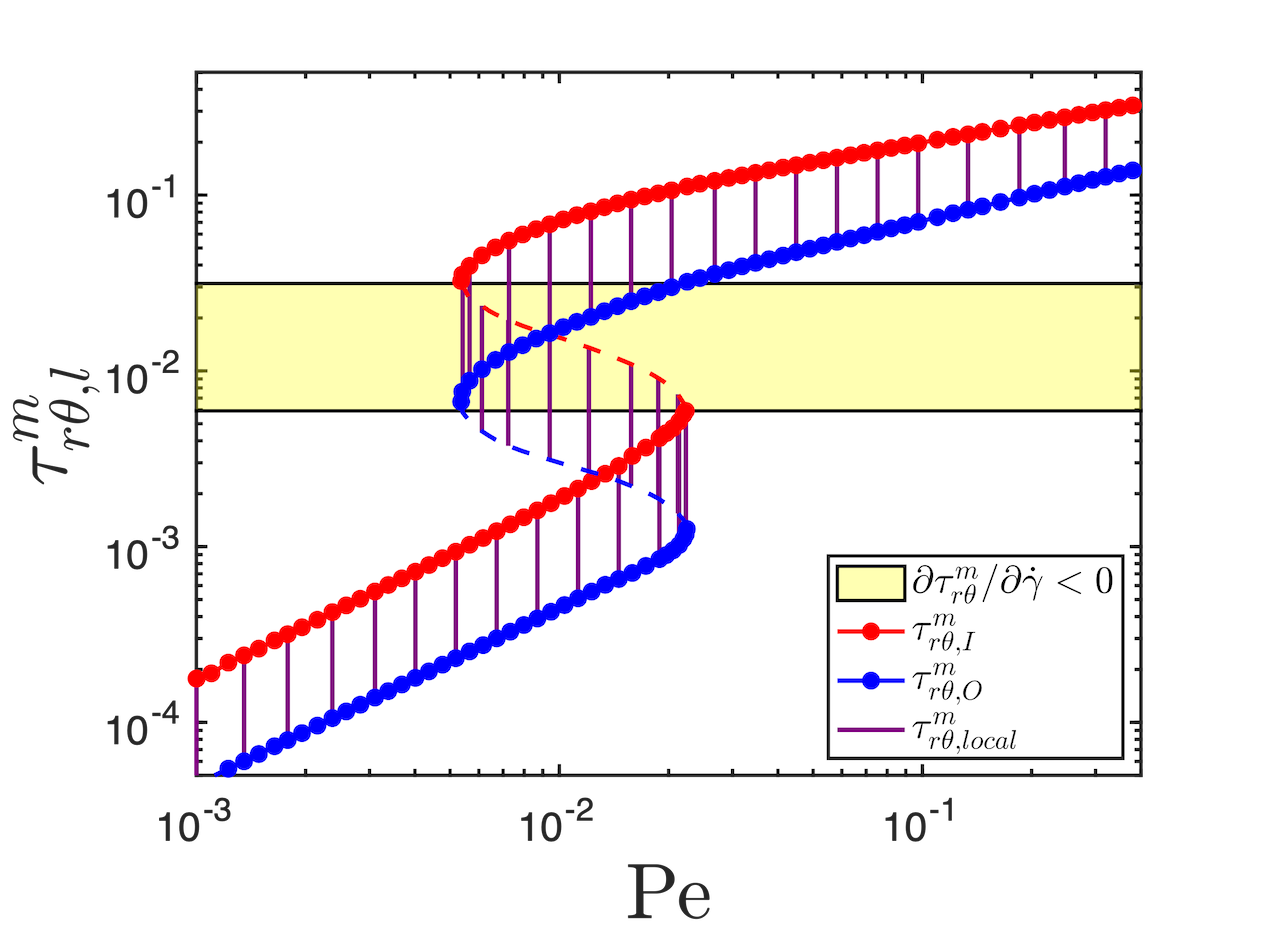}  
        \vspace{-3mm}
	    \caption{Steady state local constitutive curve for $\epsilon = 1$. The red markers are the micelle shear stress at the inner cylinder, and the blue markers are the micelle shear stress at the outer cylinder. The dashed red and blue lines represent steady states that were not explicitly calculated. The purple vertical lines connecting the markers show the local micelle shear stress throughout the gap at that Pe. The yellow rectangle shows the region where, at a given Pe, the flow will fall into the locally unstable region.}
	    \label{fig:e1_stability}
\end{figure}

Since the majority of this work considers the case of $\epsilon = 1$, it is helpful to further clarify the stability of this system at this curvature. \Cref{fig:e1_stability} shows the steady state local constitutive curve for $\epsilon = 1$. The red markers are the micelle shear stress at the inner cylinder and the blue markers are the micelle shear stress at the outer cylinder. The dashed red and blue lines represent steady states that were not explicitly calculated, but were filled in based on the underlying constitutive curve; all other steady states (shown as markers) were calculated using the Chebyshev pseudospectral code. The purple vertical lines connecting the markers show the local micelle shear stress throughout the gap at that Pe. The yellow rectangle shows the region where, at a given Pe, the flow will fall onto the locally unstable region. This plot completely describes the stability of the system for $\epsilon = 1$ and shows whether the instability will originate at the inner or the outer cylinder. \Cref{fig:e1_stability} emphasizes the features of \cref{fig:bifurcation}, but importantly it shows specifically where the instability occurs. For example, on the upper branch we find that for $\mathrm{Pe} \lesssim 0.022$ the micelle shear stress at the outer cylinder (blue markers) begins to enter the unstable region, indicating that any instability arising in the flow will likely originate close to the outer cylinder.

\subsection{Finger-like instabilities in 2D}
\label{sec:Finger_2D}
As has been discussed extensively, regions of the constitutive curve with a negative shear stress vs. shear rate slope are unstable in nature. Up until now, we have thoroughly investigated the stability of the constitutive curve and how it relates to the geometry curvature, but we have not yet investigated how this instability actually manifests. We now wish to characterize the nature of this instability in dilute WLM solutions using CFD simulations to evolve the governing equations in time. In particular, we wish to force the flow into unstable regions of the constitutive curve to investigate if and how the instability develops, as well as how the instability evolves. The full details of the simulations are described in \cref{sec:ComputationalMethods}.

\subsubsection{Instabilities for increasing shear rates}
\label{sec:below}
\begin{figure}
    \centering
		\vspace{-3mm}
        \includegraphics[width=0.95\linewidth]{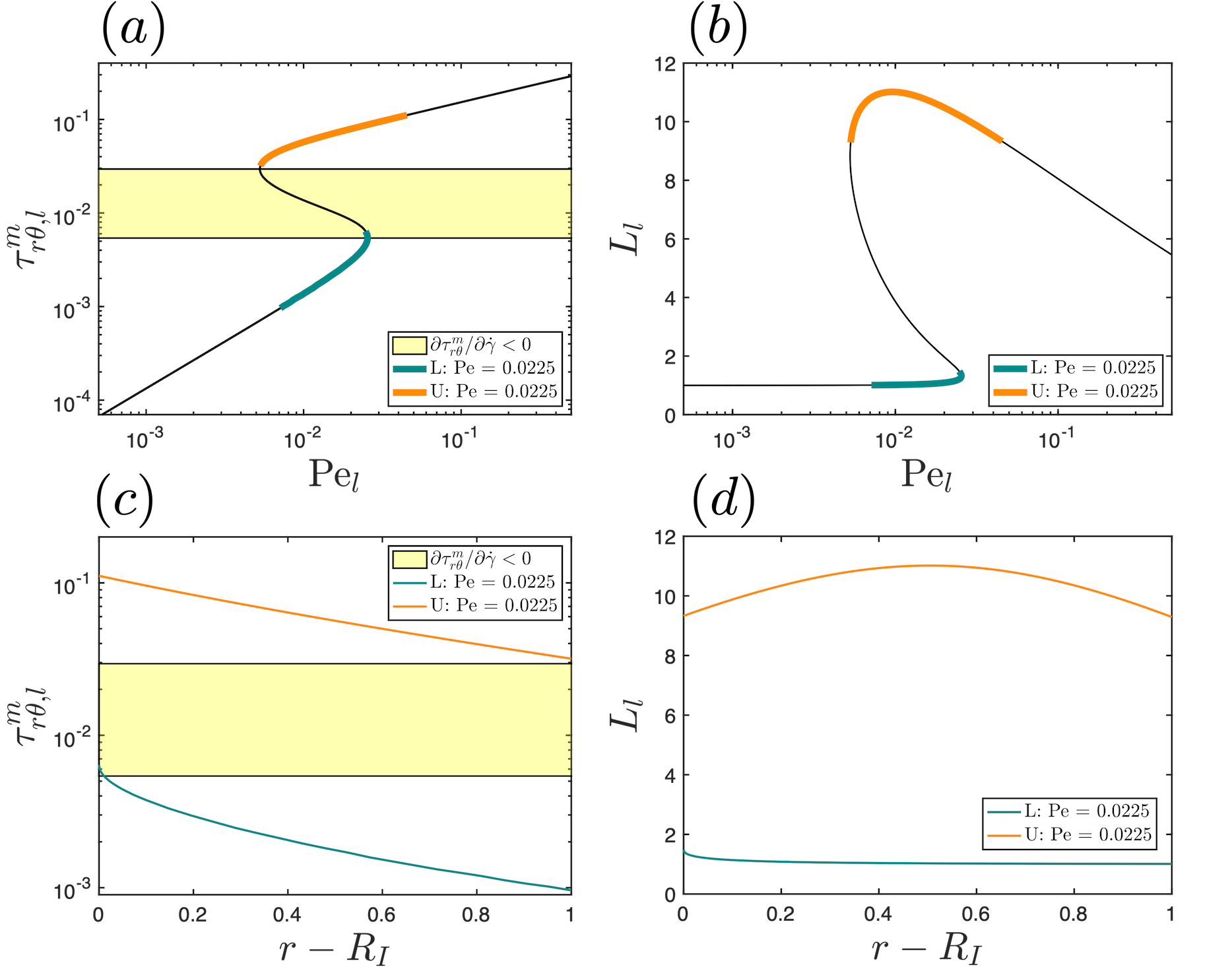}  
        \vspace{-3mm}
	    \caption{Upper and lower branch steady states for $\mathrm{Pe} = 0.0225$ at $\epsilon = 1$. (a) Local micelle shear stress and (b) local micelle length projected onto the governing constitutive curves, where the local P\'eclet number, $\mathrm{Pe}_l$, is calculated from the velocity profile throughout the gap. (c) Local micelle shear stress and (d) local micelle length over the gap radius. The orange line corresponds to the upper branch and the cyan line corresponds to the lower branch. The yellow rectangle indicates the unstable region where $\partial \tau_{r\theta}^m/\partial \dot{\gamma} < 0$.}
	    \label{fig:e1_SS_Pe0225}
\end{figure}

We begin this investigation by taking a solution at rest and slowly increasing the shear rate to $\mathrm{Pe} = 0.0225$ to force the stress near the inner cylinder into the locally unstable region. The steady states for $\mathrm{Pe} = 0.0225$ are shown in \cref{fig:e1_SS_Pe0225}, where the upper branch steady state is shown in orange and the lower branch is shown in cyan. We see from (a) and (c) that there is a small region at the inner cylinder where the micelle shear stress of the lower branch steady state begins to enter into the unstable region. The upper branch steady state, however, exists entirely on the stable upper branch; we therefore might expect that any instability will grow initially at the inner cylinder and force the flow to jump from the unstable lower branch to the stable upper branch.

\begin{figure}
    \centering
		\vspace{-3mm}
        \includegraphics[width=1\linewidth]{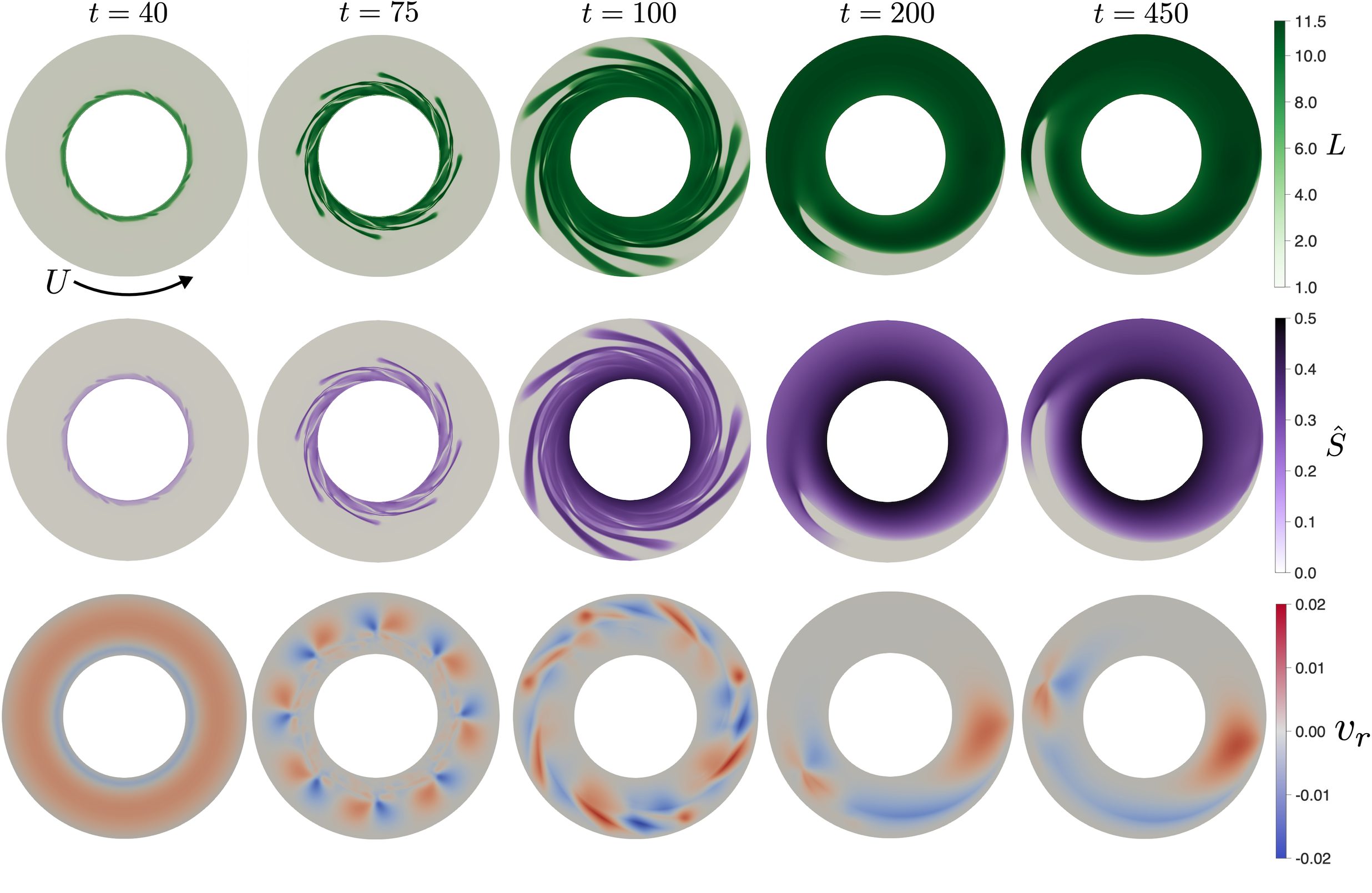}  
        \vspace{-3mm}
	    \caption{Snapshots of (top) micelle length, (middle) micelle orientation, and (bottom) radial velocity in start-up of steady shear flow with applied shear rate $\mathrm{Pe} = 0.0225$. See S1 for a movie.}
	    \label{fig:Pe0225_snapshots}
\end{figure}

\Cref{fig:Pe0225_snapshots} shows several snapshots in time of the micelle length, micelle orientation, and radial velocity in this start-up flow with an applied shear rate of $\mathrm{Pe} = 0.0225$. The SI contains a movie of this simulation. The solution is initially at rest with isotropic micelles at their equilibrium length. Upon flow inception, we see that the growth and alignment of micelles first occurs closest to the inner cylinder, where the shear stress is greatest, and then proceeds outwards through the gap. This elongation and anisotropy of micelles is associated with a radial inflow localized at the inner cylinder. At $t = 75$ we can see branches of elongated micelles growing outwards throughout the gap that are simultaneously sheared by the flow, resulting in a spiral-like pattern. The branching structures continue to grow outwards until they reach the outer cylinder ($t = 100$), at which point individual branches begin to merge into a crescent-shaped region of lower branch-micelles that rotates continuously at the outer cylinder ($t \geq 200$). In all snapshots it is clear that the elongation and anisotropy of micelles are closely intertwined. At later times we see that the sharp `interface' separating short and elongated micelles is associated with strong radial flows, which arise through the large jump in the first normal stress difference across the `interface.' We refer to this separation region as an `interface' because, although the system consists of only a single phase, the difference between regions of highly elongated and anisotropically-oriented micelles from the nearly equilibrium-length and isotropic micelles is significant enough to give the appearance of phase separation. \RJHrevise{Notably, the `interface' and resulting instability we observe closely resembles the work of Wilson and Khomami looking at interfacial instabilities between compatible polymers \cite{Wilson1993_2}. In particular, we observe similar wave- and hook-like structures, which in their study arose from shearing stresses close to the interface pulling material across the boundary into the adjacent layer. The present instability displays similar behavior, whereby elongated micelles are sheared into the region of shorter micelles.} 

\begin{figure}
    \centering
		\vspace{-3mm}
        \includegraphics[width=0.45\linewidth]{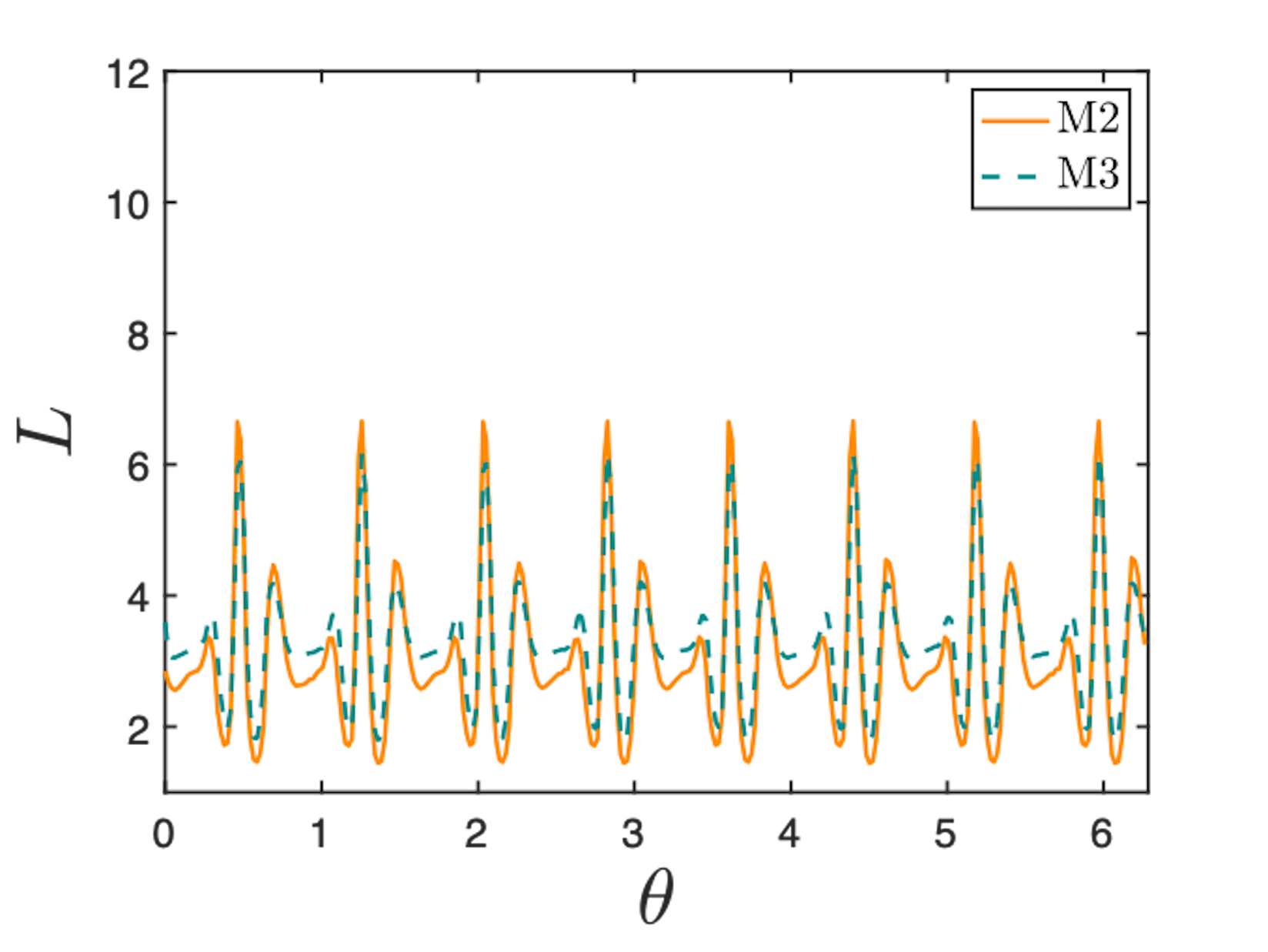}  
        \vspace{-3mm}
	    \caption{Plot of micelle length over $\theta$ at $t = 40$ and $r = R_I + 0.1d$ for start-up of steady shear flow with an applied shear rate of $\mathrm{Pe} = 0.0225$. The orange line corresponds to the M2 mesh and the cyan to the M3 mesh.}
	    \label{fig:unstableM2M3}
\end{figure}

At early times there is clearly symmetry in the flow structure that manifests as eight equally-spaced branches in $L$ and $\hat{S}$, as well as eight regions of coupled inflows and outflows. To help understand this symmetry, we consider some small perturbation to the flow of the form $\boldsymbol{B}(\boldsymbol{r}) = \boldsymbol{B}_{ss}(\boldsymbol{r}) + \widehat{\boldsymbol{B}}(\boldsymbol{r})\exp(im\theta + \sigma t)$, where $\boldsymbol{B}$ is an arbitrary flow variable (e.g., $\boldsymbol{S}$ or $L$),  $\boldsymbol{B}_{ss}$ is the steady state value of $\boldsymbol{B}$, $\widehat{\boldsymbol{B}}$ is the perturbation of $\boldsymbol{B}$, $m$ is the azimuthal wavenumber that determines the azimuthal symmetry of the instability, and $\sigma$ is a complex eigenvalue that determines the stability and growth rate of the instability. We can clearly see from \cref{fig:Pe0225_snapshots} that at $t = 40$ an unstable $m = 8$ mode appears. This mode grows radially outwards until it reaches the outer cylinder, at which point other modes set in, breaking the symmetry of the flow and leading to chaotic fluctuations. We confirmed the appearance of this $m = 8$ mode in the M1, M2, and M3 meshes; \cref{fig:unstableM2M3} shows the comparison at $r = R_I + 0.1d$ and $t = 40$ for the M2 (orange) and M3 (cyan) meshes, which show excellent agreement. We can therefore conclude that the appearance of this mode is not an artifact of the mesh resolution, and that the $m = 8$ is the most unstable mode as it pertains to this flow instability.

Returning to \cref{fig:Pe0225_snapshots}, we see that the flow has still not completely evolved to the upper branch even after 450 time units. In particular, we see a crescent-shaped region of lower branch-micelles that rotates continuously near the outer cylinder. This flow state is interesting because it demonstrates the coexistence of complex and simple states: the complex state being this rotating crescent and the simple state being a flow that has fully evolved away from this unstable state and saturated onto the stable upper branch. It is possible that this flow state will eventually evolve to the stable upper branch if given sufficient time. Consider \cref{fig:bifurcation}, which shows that the $(\epsilon,\mathrm{Pe}) = (1,0.0225)$ applied here is extremely close to the boundary of P\'eclet values that are always unstable, and therefore it is possible that the proximity to this unstable region is impeding the evolution of the flow to the stable upper branch. 

\begin{figure}
    \centering
		\vspace{-3mm}
        \includegraphics[width=1\linewidth]{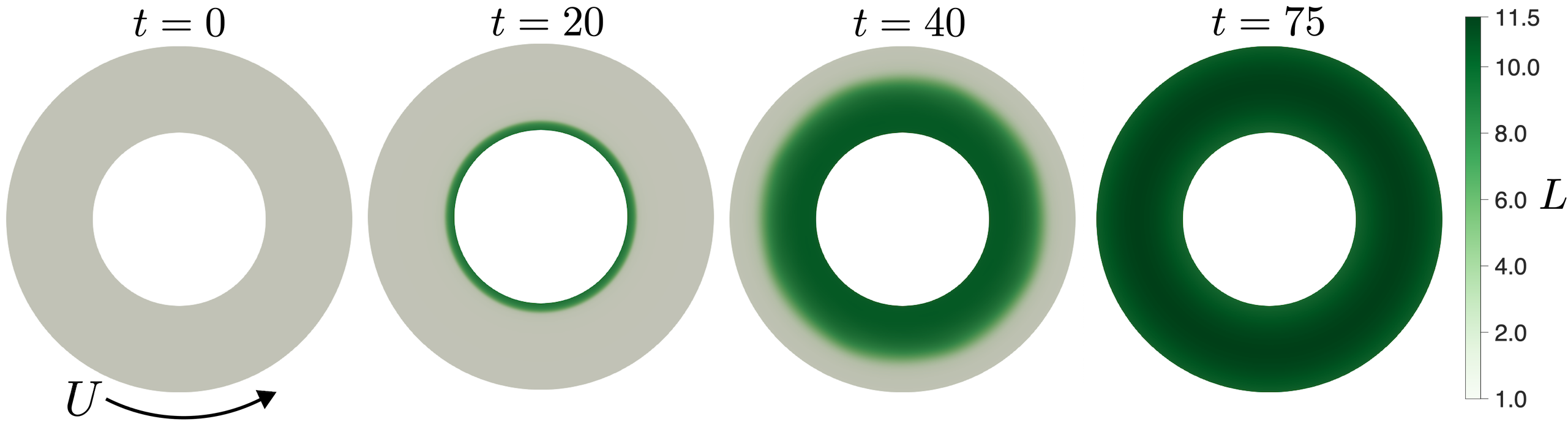}  
        \vspace{-3mm}
	    \caption{Snapshots of micelle length in start-up of steady shear flow with applied shear rate $\mathrm{Pe} = 0.024$.}
	    \label{fig:Pe024_snapshots}
\end{figure}

\begin{figure}
    \centering
		\vspace{-3mm}
        \includegraphics[width=0.5\linewidth]{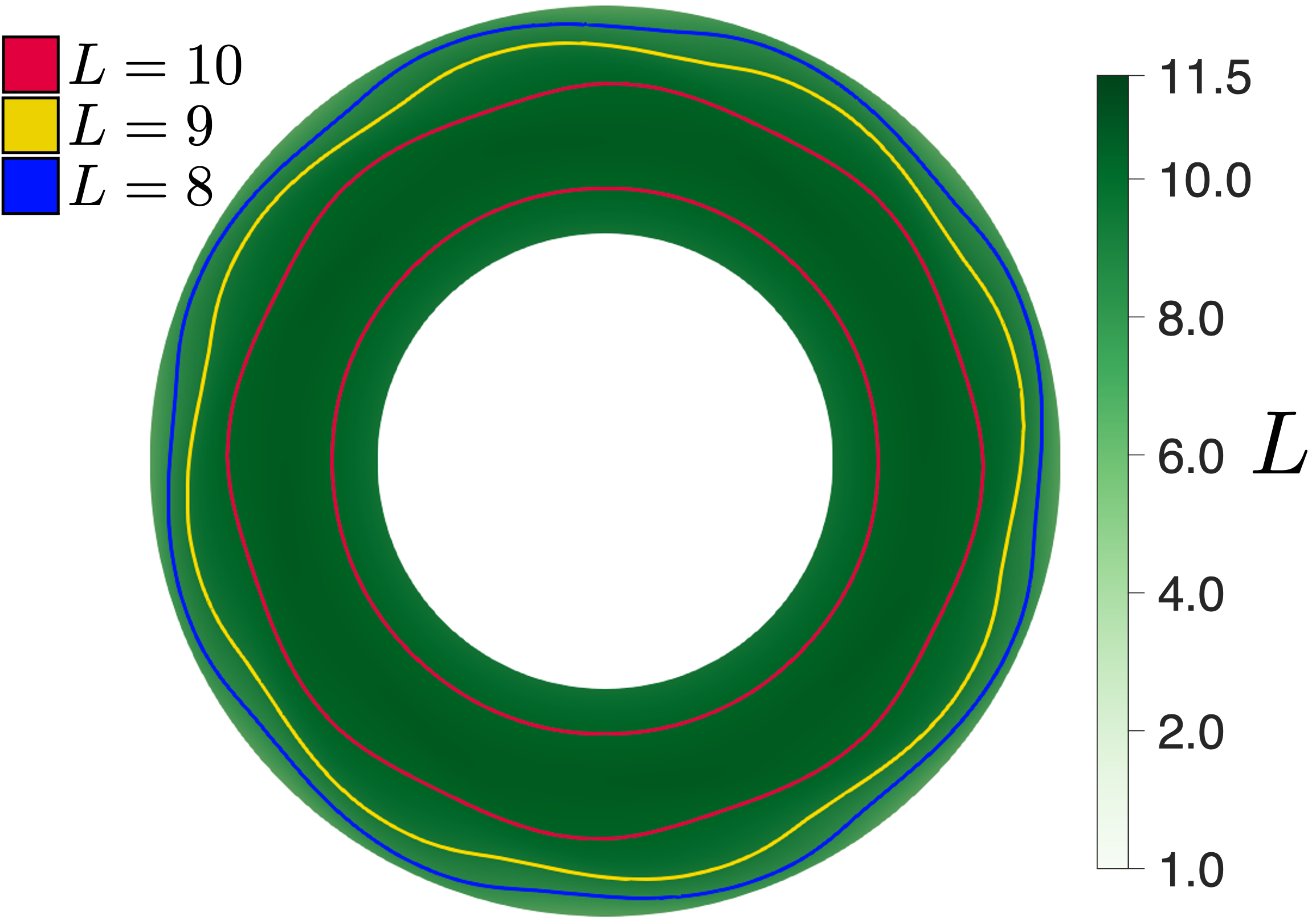}  
        \vspace{-3mm}
	    \caption{Snapshot of micelle length with contours at blue: $L = 8$, yellow: $L = 9$, and red: $L = 10$ for start-up of steady shear flow with $\mathrm{Pe} = 0.024$.}
	    \label{fig:Pe024_t45}
\end{figure}

\MDGrevise{Indeed, by increasing to $\mathrm{Pe} = 0.024$, so that the upper branch is further from the unstable region, we observe very different behavior.} \Cref{fig:Pe024_snapshots} shows several snapshots of the micelle length in this start-up flow. As before, micelle elongation occurs first at the inner cylinder and then extends outwards throughout the gap. At this larger shear rate, however, the branch structures that developed for $\mathrm{Pe} = 0.0225$ are no longer evident as the flow evolves too quickly to the stable upper branch for these branches to materialize. However, we can actually see the faint traces of these structures as well as the $m = 8$ mode in \cref{fig:Pe024_t45}. This figure shows a snapshot of micelle length with contours at blue: $L = 8$, yellow: $L = 9$, and red: $L = 10$, all at $t = 45$. The azimuthal contours clearly show the $m = 8$ mode that was previously observed, but now the mode is obscured by the rapid evolution to the upper branch. The outer contour of $L = 10$ is related to the growth of the initial instability and displays the wave-like structure of the $m = 8$ mode, as do the $L = 8$ and $L = 9$ contours; the inner $L = 10$ contour, on the other hand, is due to micelles that have begun saturating on the upper branch, evidenced by the $\theta$-independent profile. By $t = 50$ the flow has fully reached steady state, suggesting that the persistent azimuthally inhomogeneous flow observed for $\mathrm{Pe} = 0.0225$ is indeed related to the proximity to the unstable region.

\subsubsection{Instabilities for decreasing shear rates}
\label{sec:above}
\begin{figure}
    \centering
		\vspace{-3mm}
        \includegraphics[width=0.95\linewidth]{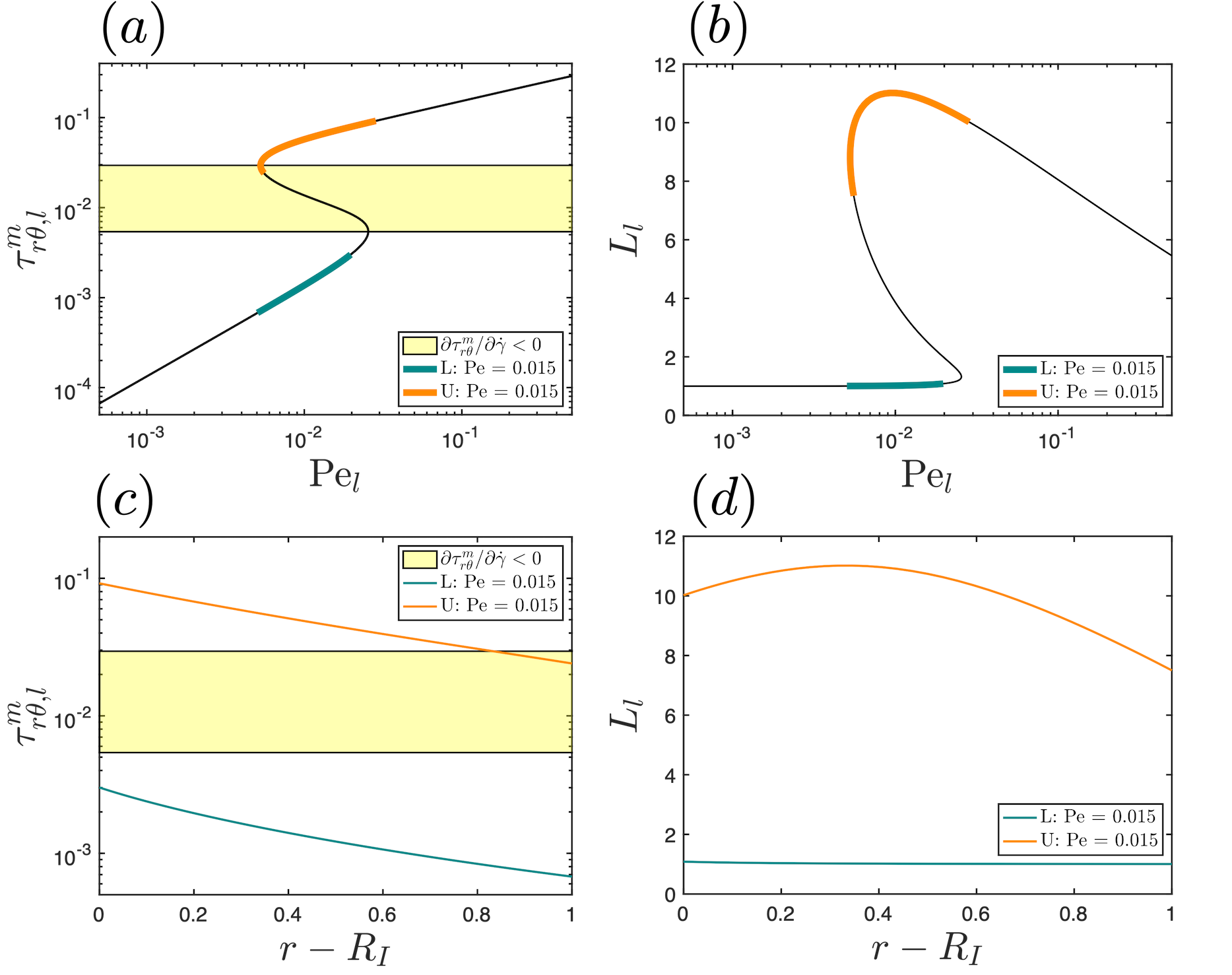}  
        \vspace{-3mm}
	    \caption{Upper and lower branch steady states for $\mathrm{Pe} = 0.015$ at $\epsilon = 1$. (a) Local micelle shear stress and (b) local micelle length projected onto the governing constitutive curves, where the local P\'eclet number, $\mathrm{Pe}_l$, is calculated from the velocity profile throughout the gap. (c) Local micelle shear stress and (d) local micelle length over the gap radius. The orange line corresponds to the upper branch and the cyan line corresponds to the lower branch. The yellow rectangle indicates the unstable region where $\partial \tau_{r\theta}^m/\partial \dot{\gamma} < 0$.}
	    \label{fig:Pe015_steadyStates}
\end{figure}

Up until now, the investigation has focused on instabilities that arise from increasing $\mathrm{Pe}$ so that the stress at the \textit{inner} cylinder enters into the unstable region; we now wish to characterize the instabilities that develop upon decreasing $\mathrm{Pe}$ so that the stress at the \textit{outer} cylinder falls into the unstable region. We expect that instabilities will now occur towards the outer cylinder. We begin by taking a steady state at $\mathrm{Pe} = 0.04$, corresponding to flow that is on the stable upper branch, and then dropping the applied P\'eclet number to $\mathrm{Pe} = 0.015$ so that the stress at the outer cylinder falls into the unstable middle branch region. The steady states for $\mathrm{Pe} = 0.015$ are shown in \cref{fig:Pe015_steadyStates}, where the upper branch steady state is shown in orange and the lower branch is shown in cyan. We see from (a) and (c) that for the upper branch steady state there is a small region at the outer cylinder where the micelle shear stress begins to enter into the unstable region. The lower branch steady state, however, is entirely stable.

\begin{figure}
    \centering
		\vspace{-3mm}
        \includegraphics[width=1\linewidth]{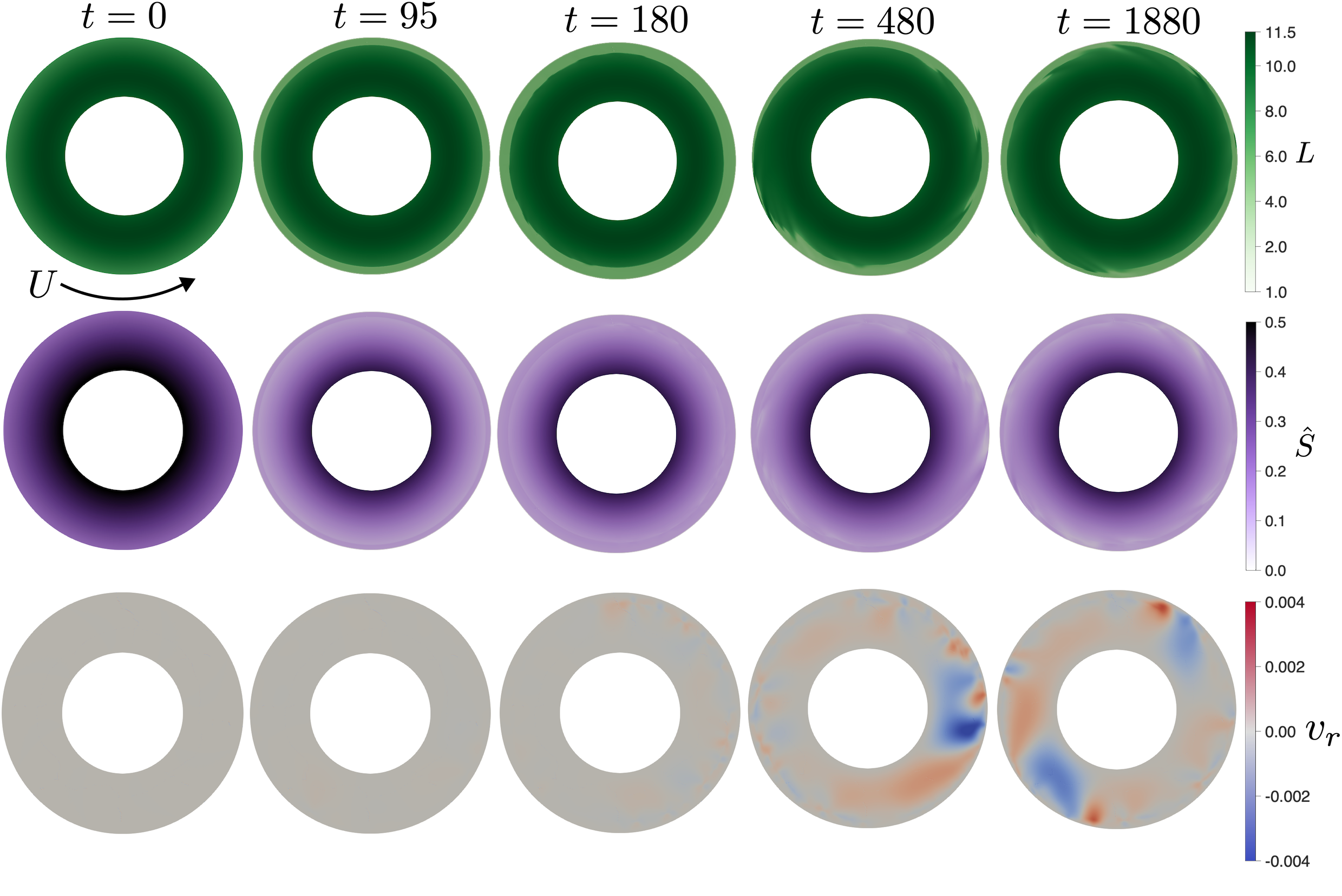}  
        \vspace{-3mm}
	    \caption{Snapshots of (top) micelle length, (middle) micelle orientation, and (bottom) radial velocity for decreasing shear flow from an applied shear rate of $\mathrm{Pe} = 0.04$ to $\mathrm{Pe} = 0.015$. See S2 for a movie.}
	    \label{fig:Pe015_snapshots}
\end{figure}

\Cref{fig:Pe015_snapshots} shows several snapshots of the micelle length, micelle orientation, and radial velocity when decreasing from an applied shear rate of $\mathrm{Pe} = 0.04$ to $\mathrm{Pe} = 0.015$. At $t = 0$ we see a nearly homogeneous flow with all micelle lengths above about $L \geq 9$, corresponding to the $\mathrm{Pe} = 0.04$ steady state. This flow slowly gives way over the next $\sim 95$ time units to a profile that exhibits a banded profile along the gradient direction. This banding is very obvious in the length profile but shows up only weakly in the orientation profile; banding at this stage does not appear in the velocity field, indicating that we are not observing traditional gradient banding. At $t = 95$ the micelle length displays an `interface'-like boundary that splits the flow into two distinct regions. Again, this is not a true interface in the traditional phase separation sense since the flow here is single phase; instead, the `interface'-like profile we are observing is strictly due to extremely sharp gradients in micelle length. As was discussed above, however, the difference between the two regions is quite significant and gives the appearance of phase separation.

At $t = 180$ we observe that the `interface'-like profile, which had previously been axisymmetric, has begun to destabilize in the form of non-axisymmetric wisps that rotate with the fluid velocity. These wisps originate in the region of elongated micelles near the inner cylinder and spread outwards throughout the gap. We also begin to see the development of a weak radial flow as the `interface' destabilizes. This destabilization process continues and grows for several hundred time units until the wispy streaks have fully extended to the outer cylinder ($t = 480$). The resulting chaotic flow state continues to fluctuate with wisps repeatedly growing and retracting, very much resembling the process described by Liu and Pine \cite{Liu1996}. At some points ($t = 1880$) the streaks almost fully encompass the gap, but at no point does the flow ever reach a steady state. The fluctuations of these wisps, and the resulting stress gradients, induce radial flows throughout the domain. Similar to the previous case of $\mathrm{Pe} = 0.0225$, this persistent chaotic flow state demonstrates the coexistence of simple and complex flow states; again, the complex flow state is observed here and the simple state is given by the stable lower branch steady state. Considering \cref{fig:bifurcation}, $\mathrm{Pe} = 0.015$ is clearly outside the region of P\'eclet numbers that are always unstable, suggesting that the persistent fluctuations do not result from close proximity to the unstable region. To help us understand why this chaotic flow state does not give way to a steady state we decreased the P\'eclet number even further, to $\mathrm{Pe} = 0.01$, to see if this lower shear rate would push the flow to the stable lower branch.

\begin{figure}
    \centering
		\vspace{-3mm}
        \includegraphics[width=1\linewidth]{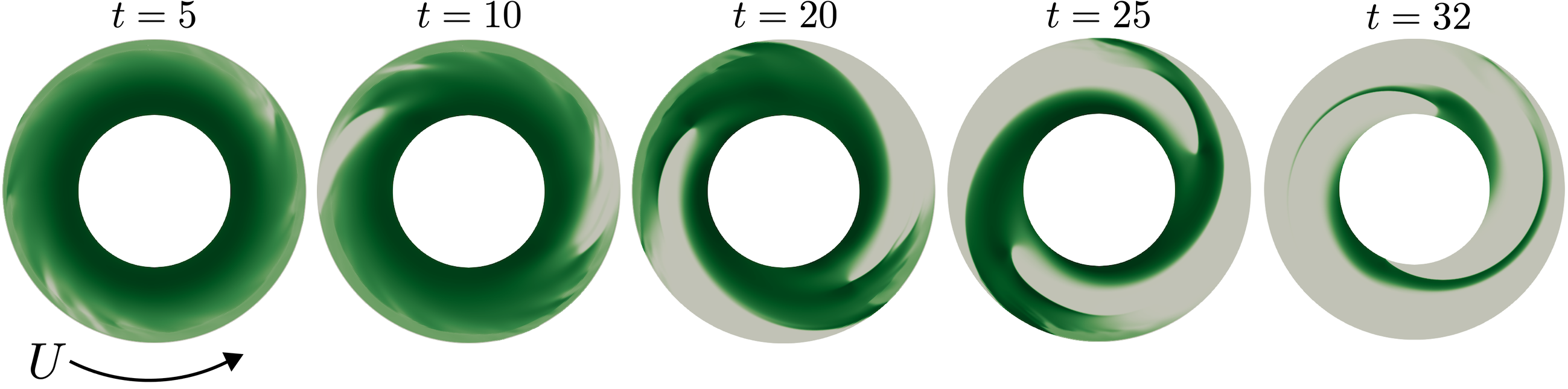}  
        \vspace{-3mm}
	    \caption{Snapshots of micelle length for decreasing shear flow from an applied shear rate of $\mathrm{Pe} = 0.015$ to $\mathrm{Pe} = 0.01$}
	    \label{fig:Pe01_snapshots}
\end{figure}

\begin{figure}
    \centering
		\vspace{-3mm}
        \includegraphics[width=1\linewidth]{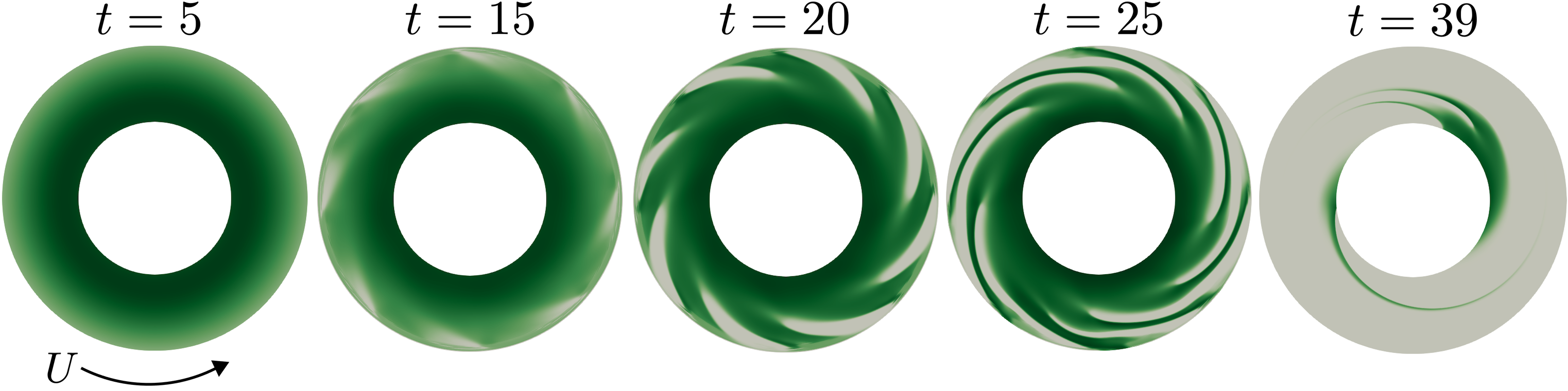}  
        \vspace{-3mm}
	    \caption{Snapshots of micelle length for decreasing shear flow from an applied shear rate of $\mathrm{Pe} = 0.025$ to $\mathrm{Pe} = 0.01$.}
	    \label{fig:Pe01_snapshots_2}
\end{figure}

Indeed, upon decreasing the applied P\'eclet number further to $\mathrm{Pe} = 0.01$ we observe that the chaotic flow state subsides and the flow rapidly approaches the stable lower branch. \Cref{fig:Pe01_snapshots} shows several snapshots of micelle length after decreasing $\mathrm{Pe}$, where we have defined $t = 0$ as the time when the reduced shear rate is implemented. In both $t = 5$ and $t = 10$ we see remnants of the `interface'-like region amidst the elongated micelle streaks. As time progresses, the `interface'-like region is no longer apparent and the flow splits into two `wings' of elongated micelles that gradually thin as they rotate. This decay to the lower branch appears to follow some $m = 2$ dependence that is obscured by other modes present in the chaotic flow state. We expect that this flow will evolve to the lower branch, but the extremely sharp gradients here led to numerical difficulties that prevented us from continuing this simulation. \RJHrevise{We also investigated dropping a steady state on the upper branch directly to $\mathrm{Pe} = 0.01$, which could potentially avoid any chaotic fluctuations that would obscure specific modes that develop in the decay to the lower branch. \Cref{fig:Pe01_snapshots_2} shows several snapshots of micelle length for a drop from $\mathrm{Pe} = 0.025$ to $\mathrm{Pe} = 0.01$; we chose $\mathrm{Pe} = 0.025$ as the initial condition rather than $\mathrm{Pe} = 0.04$ so that the reduction in shear rate is less abrupt. In general, we observe very similar behavior to the previous case, but now at $t = 15$ we observe a structure developing that resembles an eight-spoke wheel; this structure becomes further resolved at $t = 20$, developing into a spiral-like structure with eight distinct branches. Once again it seems that the $m = 8$ mode is present in the development of this instability. The fact that the drop to $\mathrm{Pe} = 0.01$ caused the flow to settle onto the stable lower branch but the drop to $\mathrm{Pe} = 0.015$ led to chaotic fluctuations suggests that the extent or amount of the flow in the locally unstable region plays a role in the final dynamics of the instability.} 

\subsection{Finger-like instabilities in 3D}
\label{sec:Re_3D}
In the final part of this work, we briefly investigate the structure of these finger-like instabilities in 3D. The radial and azimuthal elements of the geometry remain unchanged, but now we simulate a 3D domain with axial height $h=0.2$ and impose periodic boundary conditions at $z = \pm h/2$. In all 3D simulations we use the M1 mesh for $r\theta$, but now have twenty grid points in $z$ for a total mesh density of $1.8 \times 10^6$ points. This height and number of grid points are admittedly low, but we find that they are suitable for beginning to characterize the structure of these instabilities in 3D. \RJHrevise{In particular, we will see that the initial 3D structure that emerges has a wavelength that is much smaller than the height of the cylinder.}

\subsubsection{Instabilities for increasing shear rates}
\begin{figure}
    \centering
		\vspace{-3mm}
        \includegraphics[width=1\linewidth]{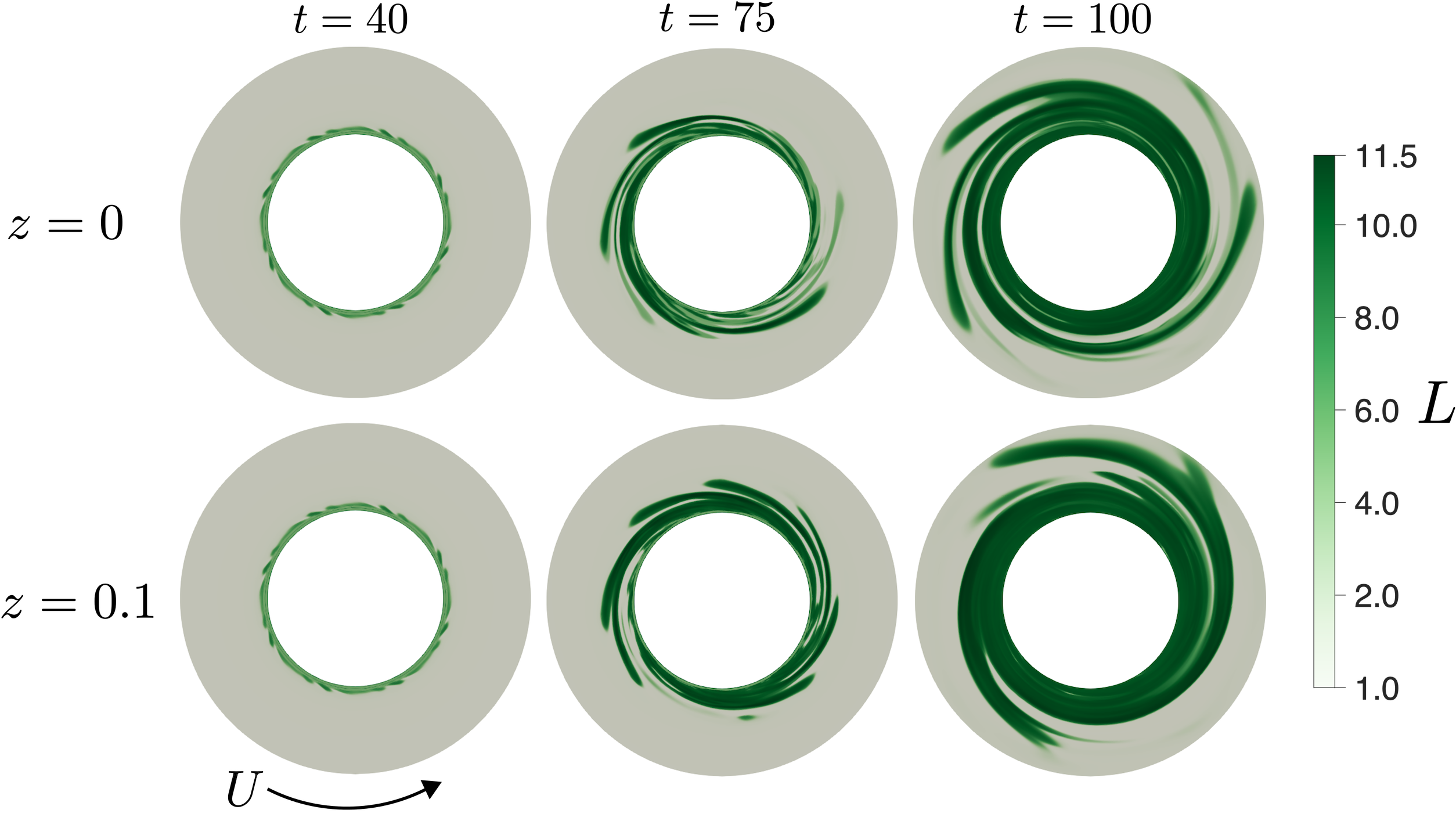}  
        \vspace{-3mm}
	    \caption{Snapshots of micelle length for increasing shear flow from rest to an applied shear rate of $\mathrm{Pe} = 0.0225$ in a 3D domain with $R_I = 1$ and $h = 0.2$. Top: $r\theta$-slice at $z = 0$ and bottom: $r\theta$-slice at $z = 0.1$. See S3 for a movie.}
	    \label{fig:3D_Pe0225}
\end{figure}

\begin{figure}
    \centering
		\vspace{-3mm}
        \includegraphics[width=1\linewidth]{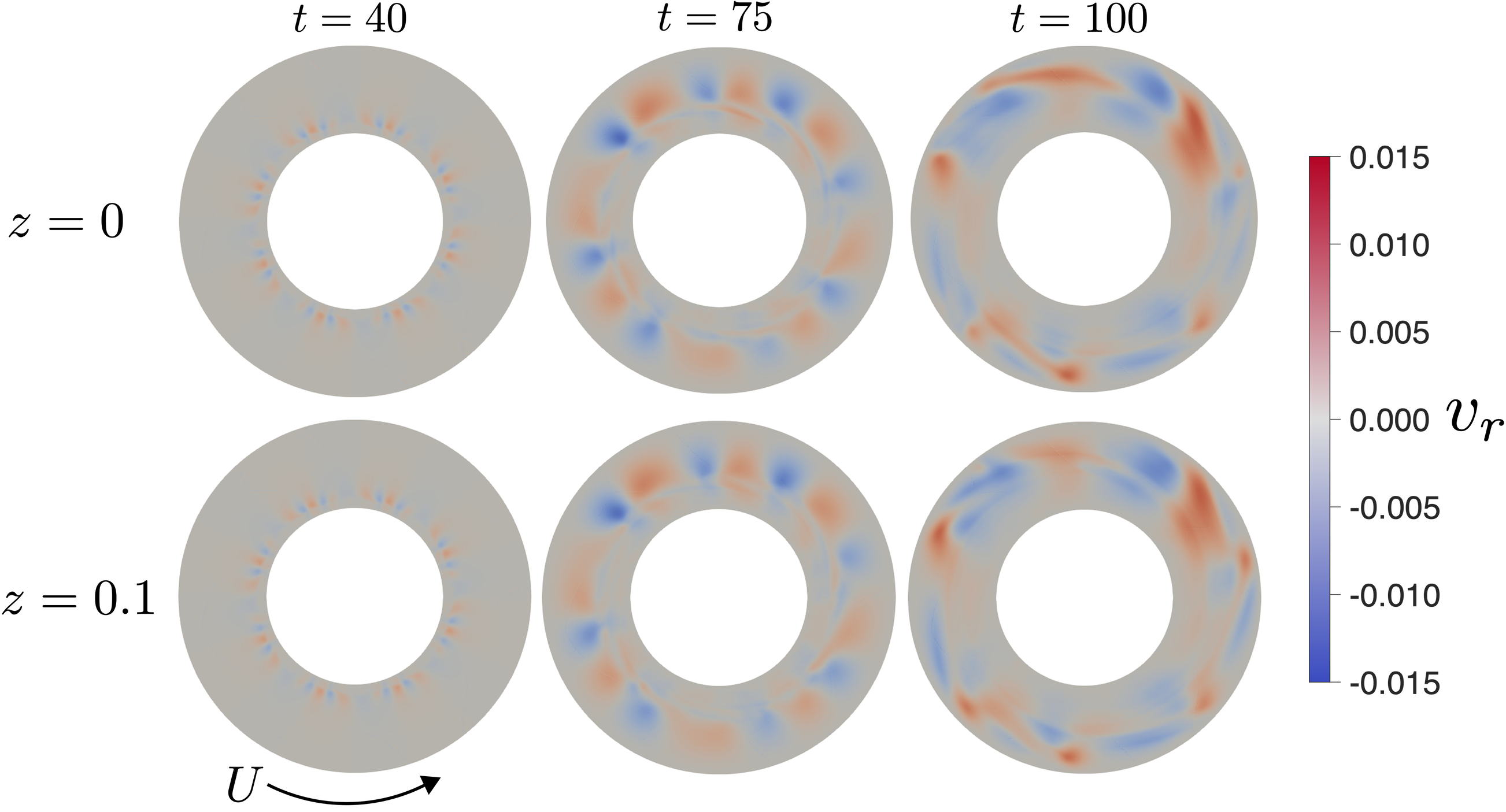}  
        \vspace{-3mm}
	    \caption{Snapshots of the radial velocity for increasing shear flow from rest to an applied shear rate of $\mathrm{Pe} = 0.0225$ in a 3D domain with $R_I = 1$ and $h = 0.2$. Top: $r\theta$-slice at $z = 0$ and bottom: $r\theta$-slice at $z = 0.1$.}
	    \label{fig:3D_Pe0225_vr}
\end{figure}

\begin{figure}
    \centering
		\vspace{-3mm}
        \includegraphics[width=0.5\linewidth]{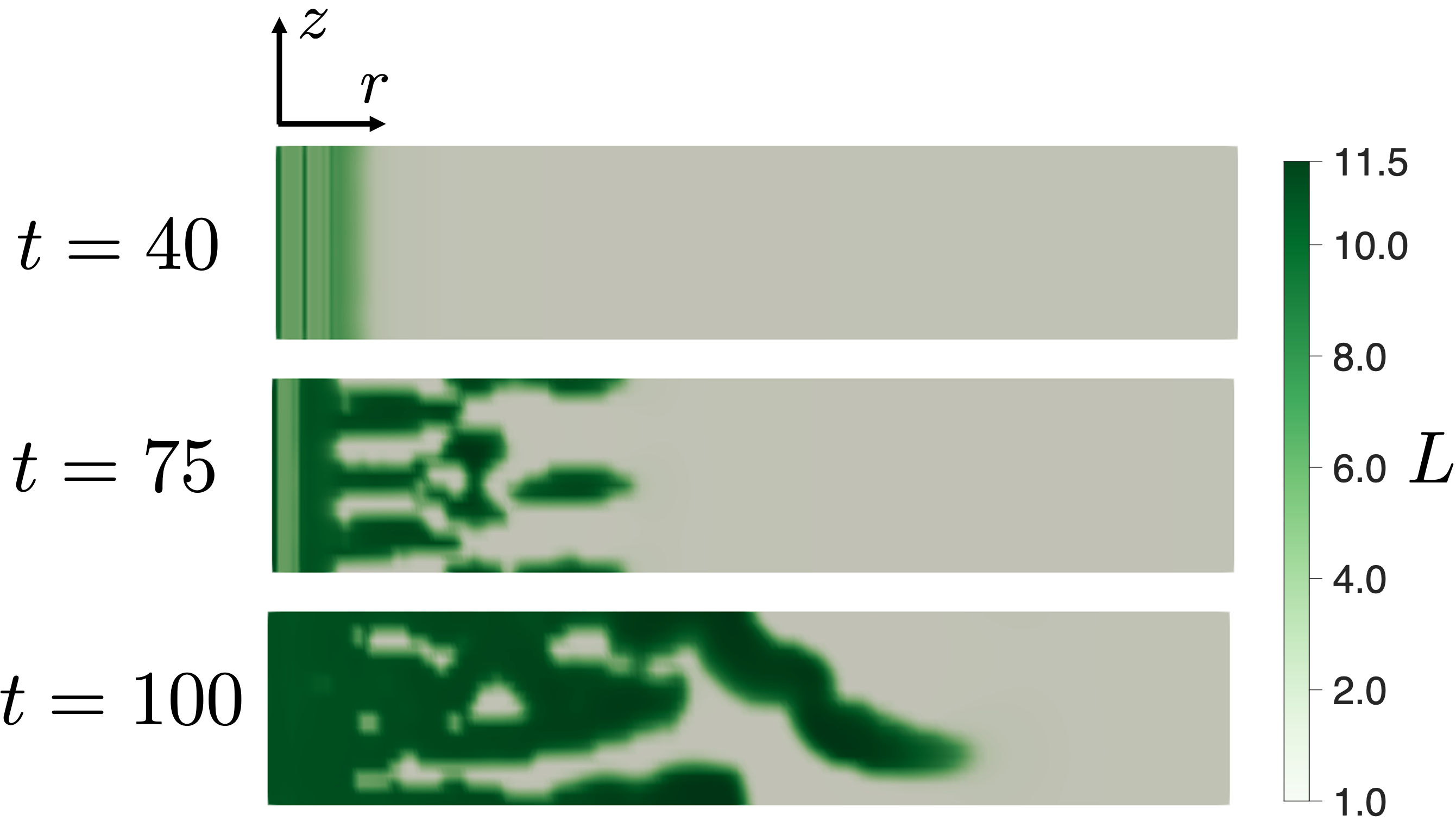}  
        \vspace{-3mm}
	    \caption{Snapshots of micelle length in the $rz$-plane at $\theta = \pi$ for increasing shear flow from rest to an applied shear rate of $\mathrm{Pe} = 0.0225$ in a 3D domain with $R_I = 1$ and $h = 0.2$. The stationary inner cylinder is on the left and the rotating outer cylinder is on the right. See S3 for a movie.}
	    \label{fig:3D_Pe0225_z}
\end{figure}

\begin{figure}
    \centering
		\vspace{-3mm}
        \includegraphics[width=0.5\linewidth]{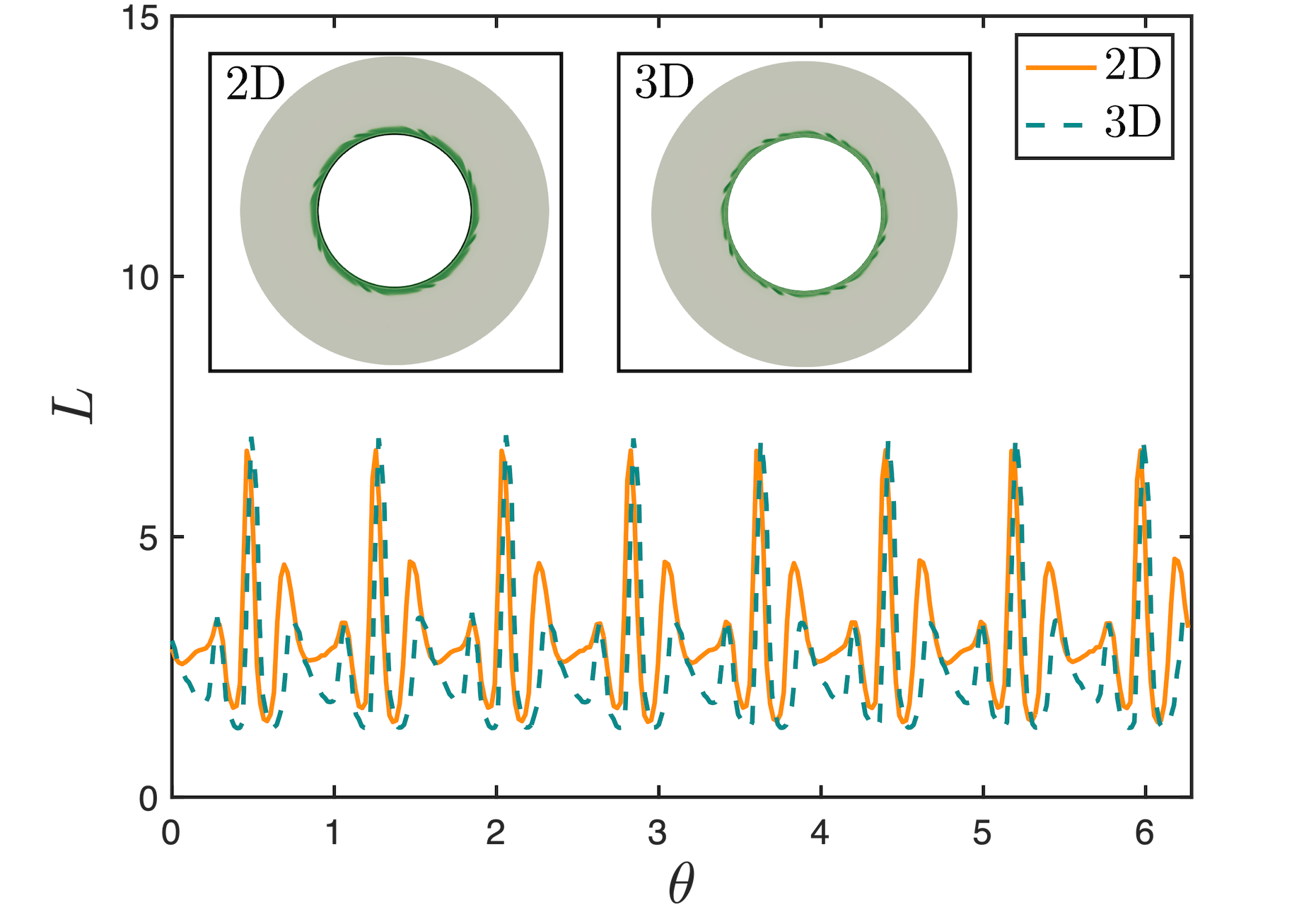}  
        \vspace{-3mm}
	    \caption{Micelle length over $\theta$ at $t = 40$ and $r = R_I + 0.1d$ for start-up of steady shear flow with an applied shear rate of $\mathrm{Pe} = 0.0225$ in 2D (orange) and 3D (cyan). The 3D slice is taken at $z = 0$, but the flow here shows no $z$-dependence.}
	    \label{fig:3D2D_mode}
\end{figure}

We begin as we did for the 2D scenarios, first by increasing the applied shear rate so that stress at the inner cylinder enters into the unstable middle branch region. Specifically, we take an isotropic solution at rest and apply a constant $\mathrm{Pe} = 0.0225$ that forces the solution into the unstable region. Again, the steady states for $\mathrm{Pe} = 0.0225$ are shown in \cref{fig:SS_Pe0225}; at steady state there will be no variation along $z$. \Cref{fig:3D_Pe0225,fig:3D_Pe0225_vr} show several $r\theta$ snapshots in time for the micelle length and radial velocity in start-up CCF with an applied shear rate of $\mathrm{Pe} = 0.0225$, where the top row corresponds the center of the cylinder, $z = 0$, and the bottom row corresponds to the top of the cylinder, $z = 0.1$. \Cref{fig:3D_Pe0225_z} shows the snapshots of micelle length but in the $rz$-plane at $\theta = \pi$, where the inner cylinder is on the left and the outer cylinder is on the right. We can see from these snapshots that up until $t = 40$ the flow is independent of $z$. In fact, at $t = 40$ the flow structure and specifically the $m = 8$ mode that appears is nearly identical to that of the 2D geometry at the same time. A direct comparison of the flow structure at $t = 40$ and $r = R_1 + 0.1d$ for 2D (orange) and 3D (cyan) is shown in \cref{fig:3D2D_mode}; we can see that the micelle length profiles exhibit the same $m = 8$ mode. This once again suggests that the $m = 8$ mode is dominant in driving the flow instability observed here, even considering axial modes, and that at $h = 0.2$ $z$-dependence does not develop until firmly after the 2D instability in the $r\theta$-plane has been established. We can conclude that the flow instability that gives rise the the finger-like structures observed here is 2D in origin, and that 3D effects are secondary to this 2D instability.

At $t = 75$ we see that the slices in \cref{fig:3D_Pe0225} now differ and the flow has developed an axial dependence. This $z$-dependence is confirmed by the variation along the vorticity axis in \cref{fig:3D_Pe0225_z}. Comparing to $t = 75$ in the 2D simulations (\cref{fig:Pe0225_snapshots}), it is clear that the structures in 3D do not retain the same symmetry that they do in 2D. This breaking of the symmetry results from the presence of axially oriented stresses and flows, whereas in 2D all dynamics are confined to the $r\theta$-plane. The variations in $z$ are much more subtle in the radial velocity snapshots, though they are present; interestingly, at $t = 75$ the 8-mode, manifested as eight coupled inflow/outflow regions, is still quite evident. 

\begin{figure}
    \centering
		\vspace{-3mm}
        \includegraphics[width=0.5\linewidth]{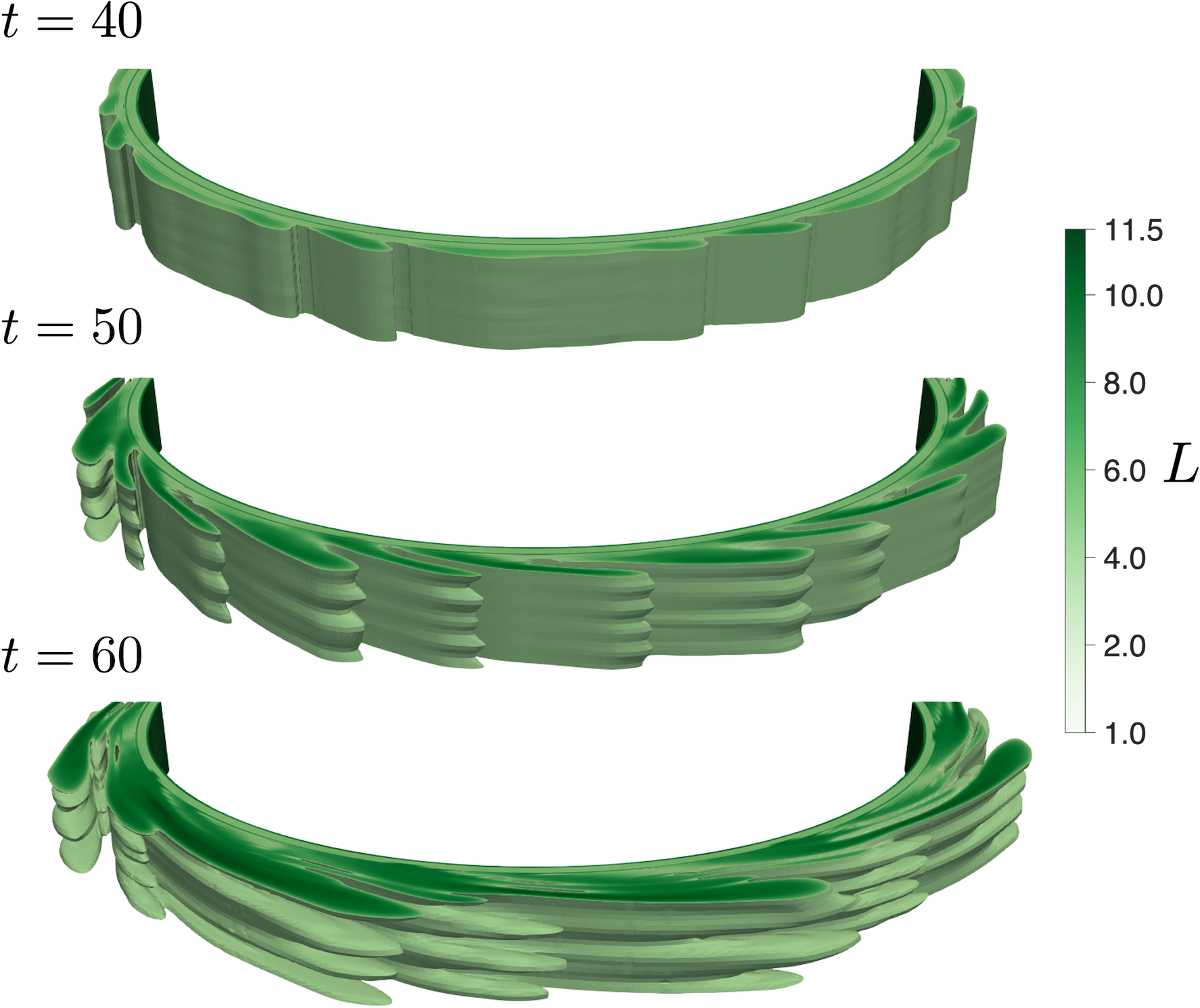}  
        \vspace{-3mm}
	    \caption{Structures of micelle length for increasing shear flow from rest to an applied shear rate of $\mathrm{Pe} = 0.0225$ in a 3D domain with $R_I = 1$ and $h = 0.2$. Top: $t = 40$, middle: $t = 50$, and bottom: $t = 60$. Structures show regions of $L \geq 5$, indicating substantial micelle elongation. See S4 for a movie.}
	    \label{fig:3D_Pe0225_isosurface}
\end{figure}

Elaborating on the development of 3D structures, \cref{fig:3D_Pe0225_isosurface} shows surfaces of elongated micelles at three different times corresponding to regions of the domain where $L \geq 5$, indicating substantially elongated micelles. The outer surface at each time is bounded by $L = 5$, but even longer micelles are present in the region close to the inner cylinder. We chose these specific times because they show the transition from 2D to 3D structures. At $t = 40$, the structures exist as 2D sheets in $r\theta$ with no variation along $z$. As discussed in \cref{sec:above}, these sheets attempt to grow outwards to the outer cylinder but are sheared azimuthally by the flow. At $t = 50$, the first variations along the vorticity axis begin to develop in the form of ripples along the sheet surfaces. These ripples appear to be somewhat sinusoidal in $z$, and although the exact wavelength is unclear, \RJHrevise{it is close to the thickness of the sheets that emerge from the 2D instability and smaller than the height of the cylinder.} Finally, at $t = 60$ the sheets have broken up along $z$ into thread-like, or finger-like, structures, signaling the transition from 2D to 3D. 

Altogether these observations emphasize that the flow instability is 2D in origin, and 3D effects arise secondarily to this 2D instability. Returning to \cref{fig:3D_Pe0225,fig:3D_Pe0225_z}, at $t = 100$ in we see that the flow has even greater variation along the vorticity axis and the finger-like structures have grown and begun to merge. Notably these finger-like structures closely resemble the finger-like structures observed in experiments. The regions of highly elongated and oriented micelles we observe here would scatter light quite differently than their isotropic and equilibrium-length counterparts, which would give rise to the variations in optics, turbidity, and birefringence observed in experiments.

\subsubsection{Instabilities for decreasing shear rates}
\begin{figure}
    \centering
		\vspace{-3mm}
        \includegraphics[width=1\linewidth]{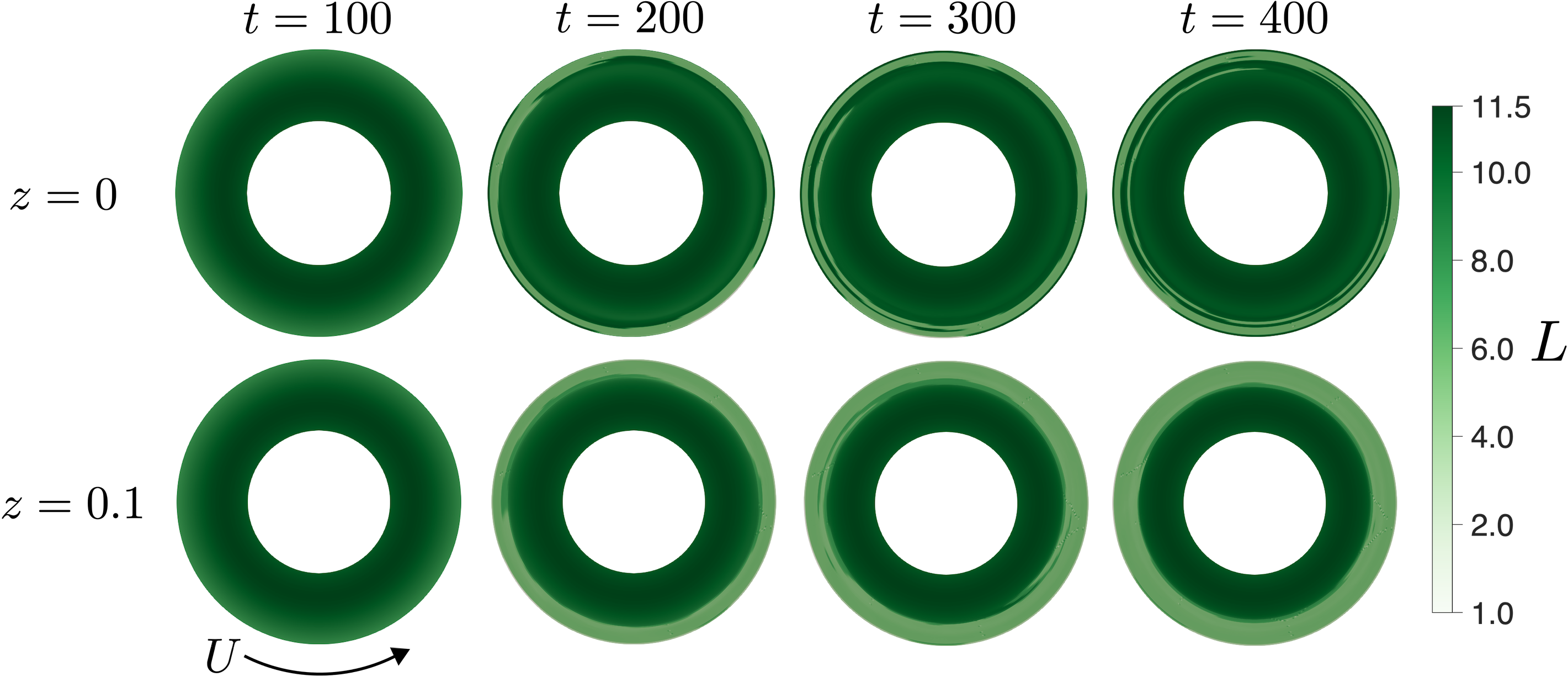}  
        \vspace{-3mm}
	    \caption{Snapshots of micelle length when decreasing the shear rate from $\mathrm{Pe} = 0.04$ to $\mathrm{Pe} = 0.015$ in a 3D domain with $R_I = 1$ and $h = 0.2$. See S5 for a movie.}
	    \label{fig:3D_Pe015}
\end{figure}

\begin{figure}
    \centering
		\vspace{-3mm}
        \includegraphics[width=0.5\linewidth]{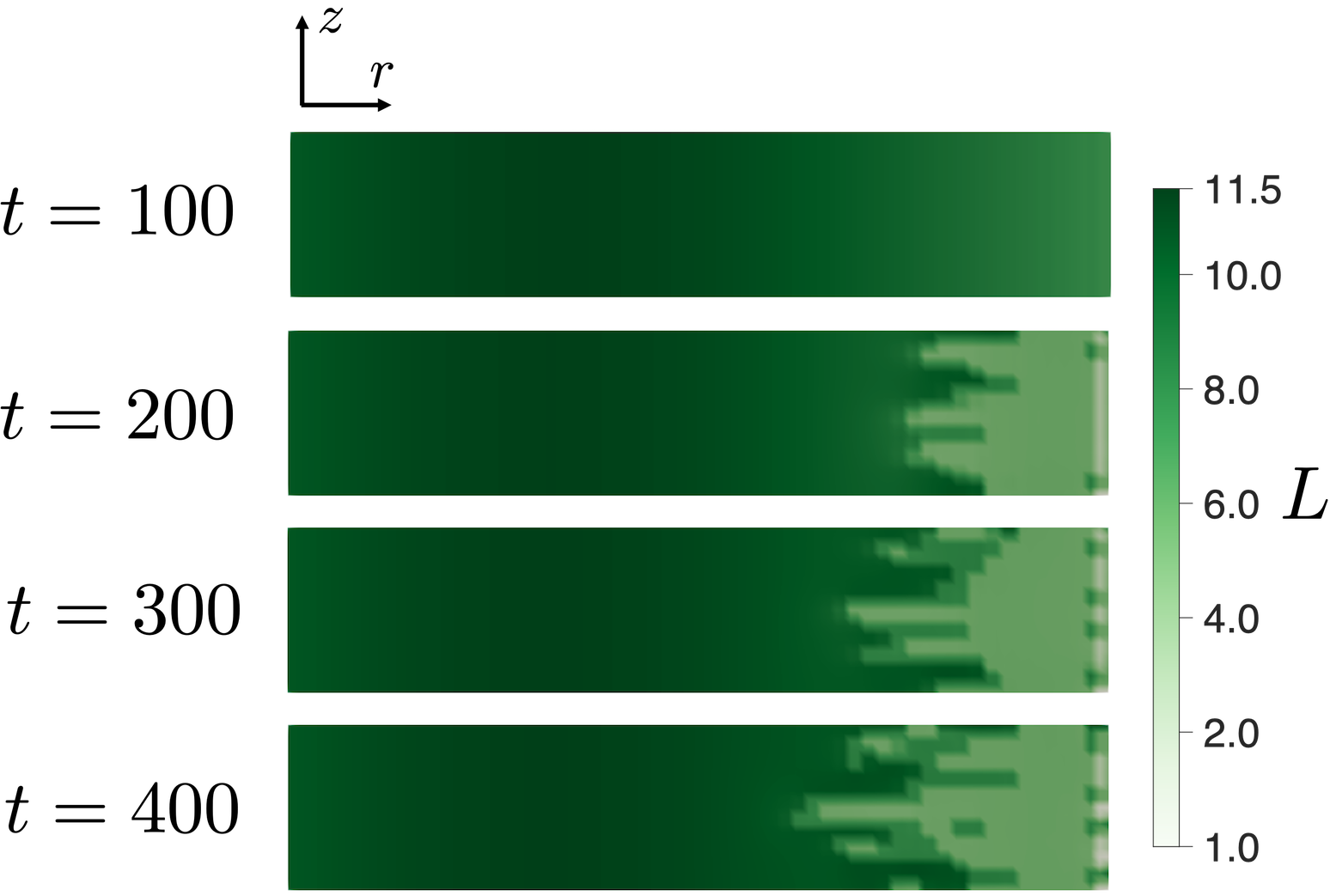}  
        \vspace{-3mm}
	    \caption{Snapshots of micelle length in the $rz$-plane at $\theta = \pi$ for decreasing the shear rate from $\mathrm{Pe} = 0.04$ to $\mathrm{Pe} = 0.015$ in a 3D domain with $R_I = 1$ and $h = 0.2$. The stationary inner cylinder is on the right and the rotating outer cylinder is on the left. See S5 for a movie.}
	    \label{fig:3D_Pe015_z}
\end{figure}

Finally, we again investigate decreasing the applied shear rate on the upper branch so that the stress at the outer cylinder falls into the unstable region, but now in 3D. We once again take a steady state at $\mathrm{Pe} = 0.04$ and decrease to $\mathrm{Pe} = 0.015$. In 2D, we observed that this reduction in shear rate transitioned the flow to a chaotic state characterized by branches of elongated micelles that extended towards the outer cylinder and were simultaneously sheared away by the flow. \Cref{fig:3D_Pe015} shows several $r\theta$ snapshots in time for the micelle length, where the top row corresponds the center of the cylinder, $z = 0$, and the bottom row corresponds to the top of the cylinder, $z = 0.1$. \Cref{fig:3D_Pe015_z} shows these snapshots but in the $rz$-plane at $\theta = \pi$, where the outer cylinder is on the left and the inner cylinder is on the right. The snapshots indicate that the behavior for this particular instability differs significantly between 2D and 3D. Comparing to \cref{fig:Pe015_snapshots}, we see that the branches of elongated micelles are much longer in 3D than 2D; in fact, at $t = 300$ we see that a single branch stretches from nearly $\theta = \pi/2$ to $\theta = 3\pi/2$. Further, there are much fewer branches in 3D than we observed for 2D. A consequence of the fewer branches is that the `interface'-like region is much more pronounced in 3D than 2D.

We also observe that there is significant variation along the vorticity axis. The $r\theta$ snapshots at both $t = 300$ and $t = 400$ for $z = 0$ show pronounced branching structures, while these same snapshots at $z = 0.1$ show minimal if any branching. The region of elongated micelles also spans further into the gap for $z = 0$. Looking at the $rz$ snapshots, we can see finger-like structures that appear fully grown around $t = 300$ and then proceed to fluctuate in time. We also are likely observing some small grid-scale artifacts in this geometry. However, we have run this exact simulation in a mesh with 50 grid points in $z$ and observed nearly identical structures. We were not able to run this simulation for as long as the 2D case, so we cannot say for certain whether this flow will eventually fall onto the lower branch where a stable steady state exists. However, we have observed that the structures shown at $t = 400$ persist with weak axial and radial fluctuations for nearly two-hundred time units, which suggests that like the 2D case the chaotic flow state here will persist indefinitely. 

To summarize this section on finger-like instabilities, we found that forcing the stress into the unstable region of the constitutive curve provoked flow structures that very closely resemble observations of finger-like structures in experiments. Notably,  the development of these structures requires a finite induction time, as in experiments, and the structures themselves are characterized by regions of elongated and anisotropic micelles that could cause the optical variations measured in experiments; similarly, these structures would produce different SANS patterns, in agreement with experimental observations. We found that the location of the instability in the gap depends on the region that first enters into the middle branch; if the stress at the inner cylinder enters into the middle branch, the instability will originate at the inner cylinder. In both increasing and decreasing the shear rate into the unstable region, the azimuthal $m = 8$ mode appears; in 3D simulations the azimuthal instability precedes variations along the vorticity axis, indicating that the flow instability is 2D in nature.

\subsection{Vorticity banding}
\label{sec:Re_VB}
\begin{figure}
    \centering
		\vspace{-3mm}
        \includegraphics[width=0.5\linewidth]{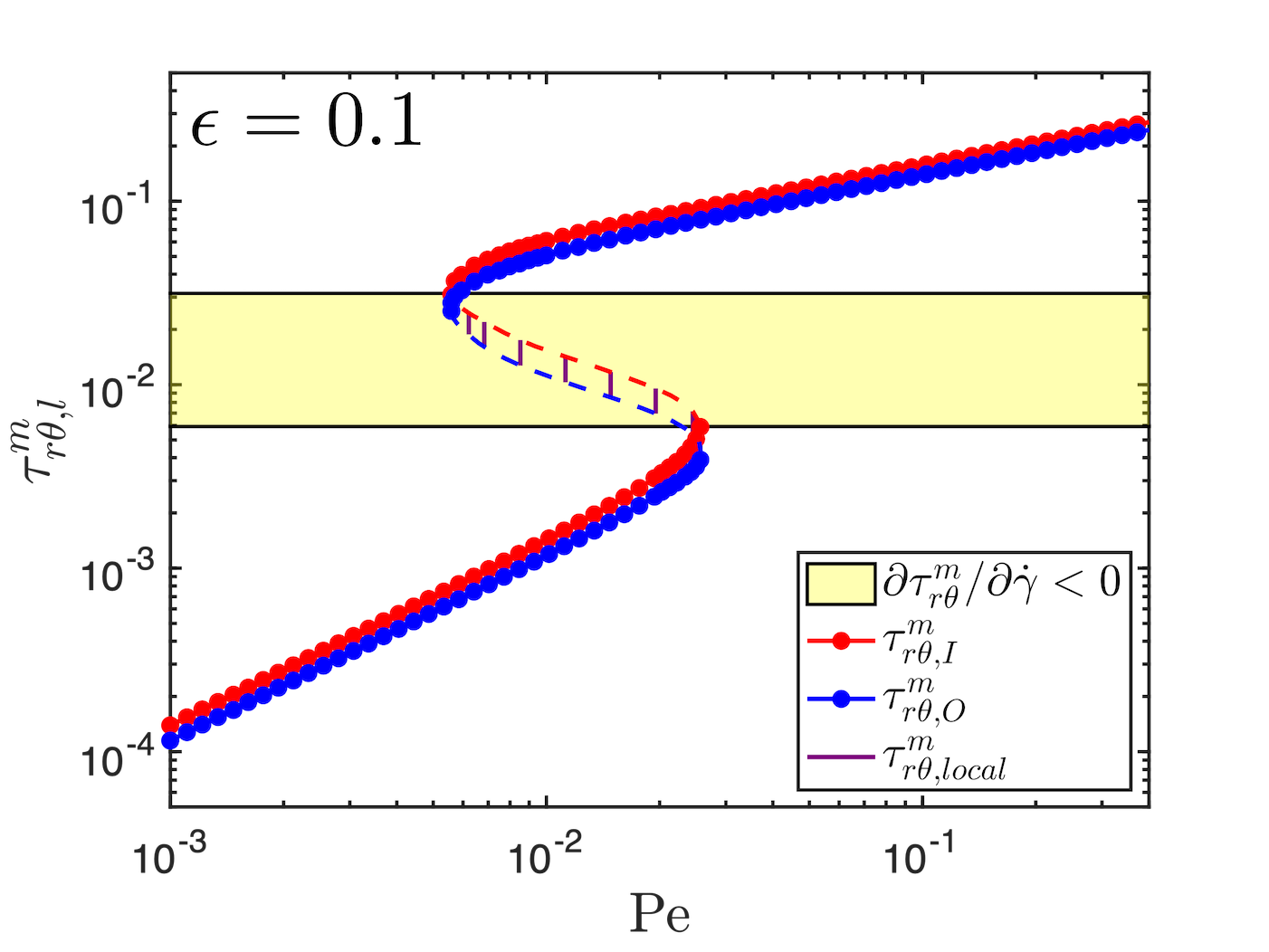}  
        \vspace{-3mm}
	    \caption{Steady state local micelle shear stress vs. applied P\'eclet number for $\epsilon = 0.1$. The red markers are the micelle shear stress at the inner cylinder, and the blue markers are the micelle shear stress at the outer cylinder. The dashed red and blue lines represent steady states that were not explicitly calculated. The purple vertical lines connecting the markers show the local micelle shear stress throughout the gap at that Pe. The yellow rectangle shows the region where, at a given Pe, the flow will fall into the locally unstable region.}
	    \label{fig:e01_stability}
\end{figure}

In the final section of this work we briefly discuss vorticity banding in CCF of dilute wormlike micelle solutions. For solutions that display a reentrant constitutive \MDGrevise{curve}, there is a region of shear rates in which two distinct steady state solutions can exist. As we saw for the $\epsilon = 1$ curve, for wide gaps it is likely that only one of the coexisting steady states will be stable because in systems with larger curvatures the range of the shear stress is often broad enough that \MDGrevise{part of the domain} falls into the unstable region of the constitutive curve and is thus susceptible to finger-like instabilities as described above. For systems with lower curvature, however, the range of the local shear stress (and other local quantities) decreases so that the formation of vorticity bands becomes more favorable. Specifically, in \cref{sec:ss} we showed that $\tau_{r\theta,O}^T = \tau_{r\theta,I}^T/(1+\epsilon)^2$ and thus as $\epsilon$ increases so does the difference between the stresses at the inner and outer cylinders. \Cref{fig:e01_stability} shows the same stability curve as \cref{fig:e1_stability}, but now for $\epsilon = 0.1$. Comparing the two, it is clear that the range of applied P\'eclet numbers that yield two stable steady states is much larger for the lower curvature geometry. For example, consider $\mathrm{Pe} = 10^{-2}$. For $\epsilon = 1$ the lower branch steady state is stable but the upper branch steady state falls into the unstable region; for $\epsilon = 0.1$, however, both lower and upper branch steady states are stable. We can use this increased stability of lower curvature systems to construct a vorticity banded solution. 

In experiments, vorticity banding typically arises in controlled-stress devices when the applied stress corresponds to the unstable region of the constitutive curve. Applying a controlled-stress boundary condition computationally requires an integral condition that we are currently unable to enforce in OpenFOAM; we are therefore limited to controlled-shear rate boundary conditions. It is also possible for vorticity banding to exist for controlled-shear rate conditions, however, an inhomogeneous flow profile will not naturally develop without some inhomogeneous perturbation or initial condition. Thus, although we cannot investigate the development of vorticity banding, we can investigate the dynamics and behavior once it manifests. 

\begin{figure}
    \centering
		\vspace{-3mm}
        \includegraphics[width=1\linewidth]{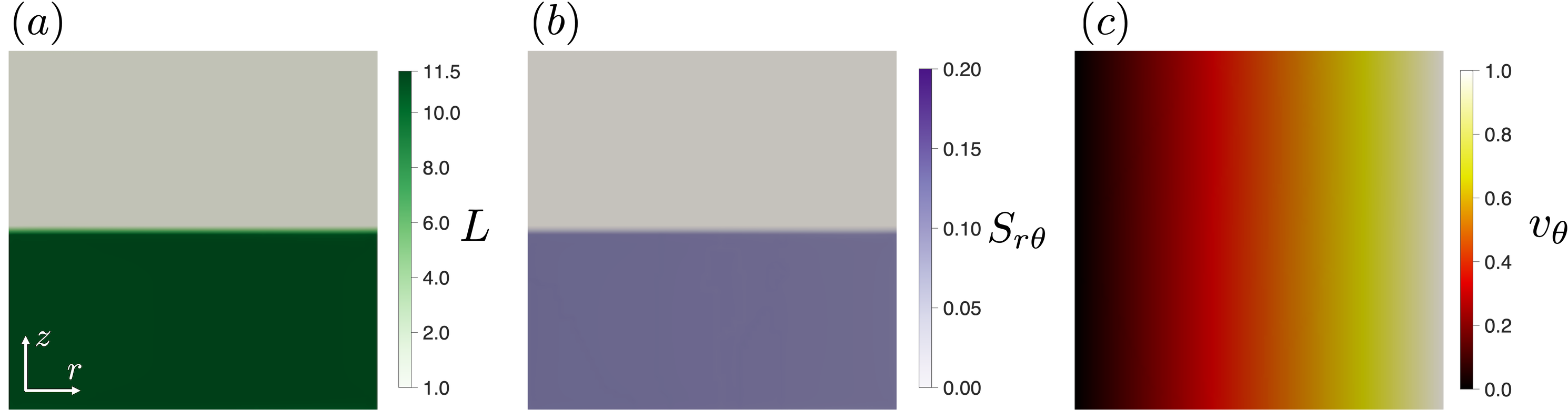}  
        \vspace{-3mm}
	    \caption{Snapshots of the final (a) length, (b) $S_{r\theta}$, and (c) azimuthal velocity for a vorticity banded initial condition. The curvature is $\epsilon =0.1$, the height is $h = 1$, and the applied shear rate is $\mathrm{Pe} = 0.01$. The rotating outer cylinder is on the right.}
	    \label{fig:vorticitySnapshots}
\end{figure}

\begin{figure}
    \centering
		\vspace{-3mm}
        \includegraphics[width=1\linewidth]{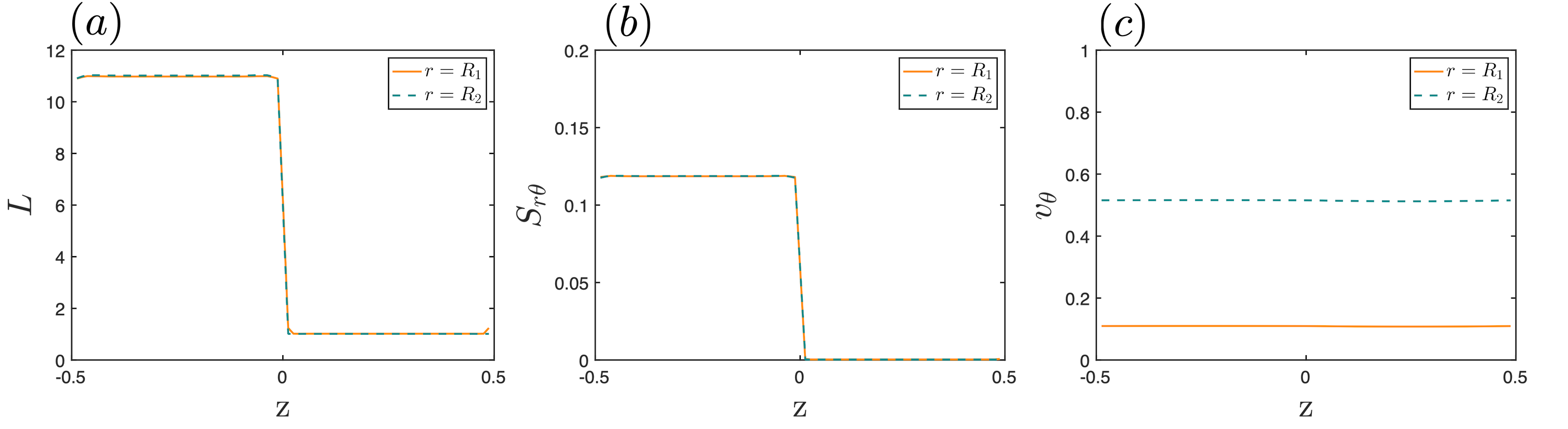}  
        \vspace{-3mm}
	    \caption{Plots of final (a) length, (b) $S_{r\theta}$, and (c) azimuthal velocity over the cylinder height at different radial locations for a vorticity banded initial condition. Orange: $r = R_I+0.1d$ and cyan: $r = R_I+0.5d$. The curvature is $\epsilon =0.1$, the height is $h = 1$, and the applied shear rate is $\mathrm{Pe} = 0.01$.}
	    \label{fig:vorticityZ}
\end{figure}

To construct a vorticity banded solution, we simply take an initial condition that is vorticity banded. Using the 3D mesh with $h = 1$ and $\epsilon = 0.1$, we construct an initial condition where half of the cylinder is on the upper branch ($z < 0$) and half is on the lower branch ($z > 0$) at an applied shear rate of $\mathrm{Pe} = 0.1$. For brevity we do not show the steady state profiles for each of these solutions, but note that at this $\mathrm{Pe}$ both steady states on the upper and lower branches are stable. We ran this vorticity banded state for a significantly long time and found that it is indeed a steady state solution. \Cref{fig:vorticitySnapshots} shows snapshots of the final (a) length, (b) $S_{r\theta}$, and (c) the azimuthal velocity where the rotating outer cylinder is on the right. \Cref{fig:vorticityZ} shows plots of these same quantities over the cylinder at two different radial locations, orange: $r = R_I+0.1d$ and cyan: $r = R_I+0.5d$. For $\epsilon = 0.1$, these quantities are nearly independent of radius. We can clearly observe the vorticity banded solution, where the length and orientation are separated along the height of the cylinder into the lower and upper branch solutions. The velocity profile, and therefore the shear rate, does not exhibit any banding because the upper and lower branches coexist at the same shear rate. This steady state profile shows that the RRM-R can capture vorticity banded solutions.

\begin{figure}
    \centering
		\vspace{-3mm}
        \includegraphics[width=1\linewidth]{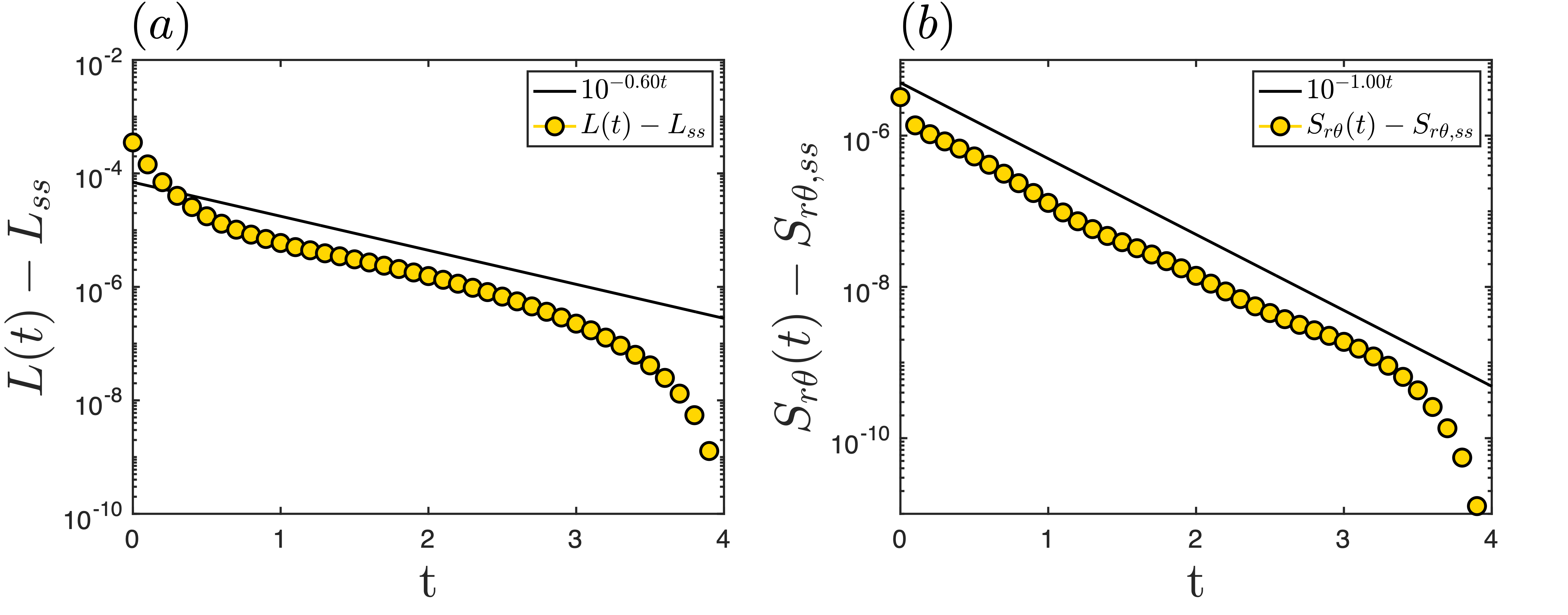}  
        \vspace{-3mm}
	    \caption{Plots of deviation from steady state values for (a) length and (b) $S_{r\theta}$ over time after some small random perturbation to a steady vorticity banded state with $\epsilon = 0.1$, $h = 1$, and $\mathrm{Pe} = 0.01$.}
	    \label{fig:VB_decay}
\end{figure}

We have also investigated the linear stability of the vorticity banded system. To accomplish this stability analysis, we applied random perturbations to the length and orientation of the micelles as well as the velocity field. In all cases we applied the perturbation to only one quantity (e.g., only the length and not the velocity or orientation), except for the case of the orientation tensor in which we randomly perturbed all components. We then tracked the perturbations over time to see if they grew, decayed, or stayed constant to determine the stability of the system. \Cref{fig:VB_decay} shows the deviation of these perturbations from steady state for (a) the length and (b) $S_{r\theta}$ over time after some small random perturbation to the vorticity banded steady state. We see that the perturbations die off very quickly and demonstrate exponential decay, indicating that vorticity banding is linear stable. The length exhibits an exponential decay with a slope of $k = -0.60$ at intermediate times. $S_{r\theta}$ shows a constant exponential decay with a slope of $k = -1.00$ for almost all times. For brevity we do not show the decay of the other orientation components or the velocity, but their decay also suggests linear stability. In future work we plan to perform rigorous linear stability analysis on this system to better characterize its stability.

It is important to note that the RRM-R does not include any translational diffusion and therefore the width of the bands as well as the location of the separation `interface' does not change with time. Again, this is a single phase system so this is not a true interface. The inclusion of translational diffusivity would likely cause the width of these bands to vary with time, though the translational diffusivity of dilute wormlike micelles is small enough, $\mathcal{O}(10^{-11} \mathrm{m}^2 \mathrm{s}^{-1})$, that the timescale of any band variation would be extraordinary long relative to other timescales in this system. Indeed, we have added diffusive terms to the micelle length and orientation and found that the evolution of the width of the bands requires an unphysically large diffusion constant. Notably, though, for stress-controlled boundary conditions the inclusion of a diffusion term may be important for fully determining the solution state \cite{Lu2000,Olmsted2008}. In future work we plan to investigate the role of diffusivity and stress-controlled boundary conditions on vorticity banded systems.

\section{Conclusions}
This study focuses on instability formation in circular Couette flow of dilute wormlike micelle solutions. We have used the reformulated reactive rod model (RRM-R), which models micelles as reactive Brownian rods, to simulate dilute WLM solutions that exhibit reentrant flow curves. We found that a reentrant flow curve, in which there is some range over which the shear stress is a multivalued function of shear rate, can provoke the development of finger-like instabilities. \RJHrevise{Specifically, we observed that the radial dependence of the shear stress in circular Couette flow allows for solutions where the domain spans both stable, $\partial \tau_{xy} /\partial \dot{\gamma} > 0$, and unstable, $\partial \tau_{xy} /\partial \dot{\gamma} < 0$, regions of the flow curve. We have found that this mixed local stability can lead to complex flow instabilities that originate in the parts of the domain existing in the unstable region of the flow curve, and that these instabilities manifest as finger-like structures comprised of elongated and anisotropically-oriented micelles.} 

Upon increasing the shear rate so that the shear stress at the inner cylinder entered into this unstable region, we found that `fingers' or branches form towards the inner cylinder and expand outwards throughout the gap. For all the increasing shear rates tested we observed the presence of an $m = 8$ azimuthal mode in the flow profiles, and this appeared in both 2D and 3D simulations. Moreover, we found that the constitutive instability is 2D in nature and 3D variations are secondary effects. In 3D simulations, this instability manifested as sheet-like structures that were constant along the vorticity axis, and as the instability grew the sheets developed surface ripples that induced breakup into thread-like, or finger-like, structures. We also observed the $m = 8$ mode when decreasing the shear rate so that the stress at the outer cylinder fell into the unstable region. When decreasing the shear rate we found that if the shear rate is reduced to a value such that only a small portion of the local stress falls into the unstable region, the flow will exhibit chaotic fluctuations that persist indefinitely despite a stable steady state existing, indicating the coexistence of simple and complex states. As the shear rate is reduced further, the system rapidly decays to the stable steady state. Finally, we found for decreasing shear rate that the branching structures observed in 3D differ significantly from those in 2D; specifically, the 3D branches are much longer and extend for nearly a radian in the azimuthal direction. 

In both cases of either decreasing or increasing the shear rate to force the stress into the unstable region, we found that the branching structures that develop very closely resemble the finger-like structures that have been observed in experiments. Moreover, these branches grow and retract in a manner that is consistent with experimental observations. These branches consist of highly elongated and oriented micelles that would scatter light differently than isotropic, equilibrium-length rods, giving rise to the various optical differences observed in experiments that show these finger-like structures. To the best of our knowledge, this study is the first computational observation of these finger-like instabilities.

We have also found that the RRM-R can capture vorticity banding, and this banding is only observed in regions where two stable steady states coexist at the same applied P\'eclet number. A consequence of this fact is that vorticity banding is more readily formed in geometries with lower curvatures since there is a broader range of applied shear rates that can give rise to two coexisting stable steady states. As the curvature of the geometry increases the multivalued stability range decreases. We have also investigated the linear stability of the vorticity banded state through the application of small, random perturbations; in all cases we found that the perturbations die off relatively quickly with some exponential decay, indicating that the vorticity banded state is linearly stable.


\section*{Acknowledgements}

This material is based on work supported by the National Science Foundation under grant number CBET-1803090 \MDGrevise{ and the Office of Naval Research under grant number  N00014-18-1-2865 (Vannevar Bush Faculty Fellowship)}.


\bibliography{mybibfile}

\begin{thebibliography}{10}
\expandafter\ifx\csname url\endcsname\relax
  \def\url#1{\texttt{#1}}\fi
\expandafter\ifx\csname urlprefix\endcsname\relax\def\urlprefix{URL }\fi
\expandafter\ifx\csname href\endcsname\relax
  \def\href#1#2{#2} \def\path#1{#1}\fi

\bibitem{Israelachvili2011}
J.~N. Israelachvili, Intermolecular and surface forces, Academic Press, 2011.

\bibitem{Oelschlaeger2010}
C.~Oelschlaeger, P.~Suwita, N.~Willenbacher, Effect of counterion binding
  efficiency on structure and dynamics of wormlike micelles, Langmuir 26~(10)
  (2010) 7045--7053.

\bibitem{Lerouge2009}
S.~Lerouge, J.-F. Berret, Shear-induced transitions and instabilities in
  surfactant wormlike micelles, in: Polymer Characterization, Springer, 2009,
  pp. 1--71.

\bibitem{Cates2006}
M.~E. Cates, S.~M. Fielding, Rheology of giant micelles, Advances in Physics
  55~(7-8) (2006) 799--879.

\bibitem{Berret1998}
J.-F. Berret, R.~Gamez-Corrales, J.~Oberdisse, L.~Walker, P.~Lindner,
  Flow-structure relationship of shear-thickening surfactant solutions, EPL
  (Europhysics Letters) 41~(6) (1998) 677.

\bibitem{Von1998}
H.~von Berlepsch, L.~Harnau, P.~Reineker, Persistence length of wormlike
  micelles from dynamic light scattering, The Journal of Physical Chemistry B
  102~(39) (1998) 7518--7522.

\bibitem{Zou2014}
W.~Zou, R.~G. Larson, A mesoscopic simulation method for predicting the
  rheology of semi-dilute wormlike micellar solutions, Journal of Rheology
  58~(3) (2014) 681--721.

\bibitem{Imae1989}
T.~Imae, The flexibility of rodlike micelles in aqueous solutions and the
  crossover concentrations among dilute, semidilute, and concentrated regimes,
  Colloid and Polymer Science 267 (1989) 707--713.

\bibitem{Ohlendorf1986}
D.~Ohlendorf, W.~Interthal, H.~Hoffmann, Surfactant systems for drag reduction:
  physico-chemical properties and rheological behaviour, Rheologica Acta 25~(5)
  (1986) 468--486.

\bibitem{Helgeson2010}
M.~E. Helgeson, T.~K. Hodgdon, E.~W. Kaler, N.~J. Wagner, A systematic study of
  equilibrium structure, thermodynamics, and rheology of aqueous ctab/nano3
  wormlike micelles, Journal of Colloid and Interface Science 349~(1) (2010)
  1--12.

\bibitem{Yang2002}
J.~Yang, Viscoelastic wormlike micelles and their applications, Current Opinion
  in Colloid \& Interface Science 7~(5-6) (2002) 276--281.

\bibitem{Zakin2017}
J.~L. Zakin, A.~J. Maxson, T.~Saeki, P.~F. Sullivan, {Turbulent Drag-reduction
  Applications of Surfactant Solutions}, in: C.~A. Dreiss, Y.~Feng (Eds.),
  Wormlike Micelles: Advances in Systems, Characterization and Applications,
  Royal Society of Chemistry, 2017, pp. 353--378.

\bibitem{Virk1975}
P.~S. Virk, Drag reduction fundamentals, AIChE Journal 21~(4) (1975) 625--656.

\bibitem{Zakin1996}
J.~L. Zakin, J.~Myska, Z.~Chara, New limiting drag reduction and velocity
  profile asymptotes for nonpolymeric additives systems, AIChE Journal 42~(12)
  (1996) 3544--3546.

\bibitem{Saeki2011}
T.~Saeki, Flow properties and heat transfer of drag-reducing surfactant
  solutions, Developments in Heat Transfer 1 (2011) 8--9.

\bibitem{Krope2010}
A.~Krope, L.~C. Lipus, Drag reducing surfactants for district heating, Applied
  Thermal Engineering 30~(8-9) (2010) 833--838.

\bibitem{Shrestha2011}
L.~K. Shrestha, M.~Yamamoto, S.~Arima, K.~Aramaki, Charge-free reverse wormlike
  micelles in nonaqueous media, Langmuir 27~(6) (2011) 2340--2348.

\bibitem{Tung2007}
S.-H. Tung, Y.-E. Huang, S.~R. Raghavan, Contrasting effects of temperature on
  the rheology of normal and reverse wormlike micelles, Langmuir 23~(2) (2007)
  372--376.

\bibitem{Keller1998}
S.~Keller, P.~Boltenhagen, D.~Pine, J.~Zasadzinski, Direct observation of
  shear-induced structures in wormlike micellar solutions by freeze-fracture
  electron microscopy, Physical Review Letters 80~(12) (1998) 2725.

\bibitem{Perge2014}
C.~Perge, M.-A. Fardin, S.~Manneville, Surfactant micelles: Model systems for
  flow instabilities of complex fluids, The European Physical Journal E 37~(4)
  (2014) 23.

\bibitem{Wu2018}
S.~Wu, H.~Mohammadigoushki, Sphere sedimentation in wormlike micelles: Effect
  of micellar relaxation spectrum and gradients in micellar extensions, Journal
  of Rheology 62~(5) (2018) 1061--1069.

\bibitem{Mohammadigoushki2017}
H.~Mohammadigoushki, S.~J. Muller, Inertio-elastic instability in
  taylor-couette flow of a model wormlike micellar system, Journal of Rheology
  61~(4) (2017) 683--696.

\bibitem{Fardin2012}
M.-A. Fardin, S.~Lerouge, Instabilities in wormlike micelle systems, The
  European Physical Journal E 35~(9) (2012) 1--29.

\bibitem{Bhardwaj2007}
A.~Bhardwaj, E.~Miller, J.~P. Rothstein, Filament stretching and capillary
  breakup extensional rheometry measurements of viscoelastic wormlike micelle
  solutions, Journal of Rheology 51~(4) (2007) 693--719.

\bibitem{Rojas2008}
M.~R. Rojas, A.~J. M{\"u}ller, A.~E. S{\'a}ez, Shear rheology and porous media
  flow of wormlike micelle solutions formed by mixtures of surfactants of
  opposite charge, Journal of Colloid and Interface Science 326~(1) (2008)
  221--226.

\bibitem{Hommel2021}
R.~J. Hommel, M.~D. Graham, Constitutive modeling of dilute wormlike micelle
  solutions: Shear-induced structure and transient dynamics, Journal of
  Non-Newtonian Fluid Mechanics 295 (2021) 104606.

\bibitem{Boltenhagen1997}
P.~Boltenhagen, Y.~Hu, E.~Matthys, D.~Pine, Observation of bulk phase
  separation and coexistence in a sheared micellar solution, Physical Review
  Letters 79~(12) (1997) 2359.

\bibitem{Herle2007}
V.~Herle, J.~Kohlbrecher, B.~Pfister, P.~Fischer, E.~J. Windhab, Alternating
  vorticity bands in a solution of wormlike micelles, Physical Review Letters
  99~(15) (2007) 158302.

\bibitem{Chen1992}
L.~Chen, C.~Zukoski, B.~Ackerson, H.~Hanley, G.~Straty, J.~Barker, C.~Glinka,
  Structural changes and orientaional order in a sheared colloidal suspension,
  Physical Review Letters 69~(4) (1992) 688.

\bibitem{Dhont2003a}
J.~K. Dhont, M.~P. Lettinga, Z.~Dogic, T.~A. Lenstra, H.~Wang, S.~Rathgeber,
  P.~Carletto, L.~Willner, H.~Frielinghaus, P.~Lindner, Shear-banding and
  microstructure of colloids in shear flow, Faraday discussions 123 (2003)
  157--172.

\bibitem{Wilkins2006}
G.~M. Wilkins, P.~D. Olmsted, Vorticity banding during the lamellar-to-onion
  transition in a lyotropic surfactant solution in shear flow, The European
  Physical Journal E 21 (2006) 133--143.

\bibitem{Caserta2012}
S.~Caserta, S.~Guido, Vorticity banding in biphasic polymer blends, Langmuir
  28~(47) (2012) 16254--16262.

\bibitem{Mutze2014}
A.~M{\"u}tze, P.~Heunemann, P.~Fischer, On the appearance of vorticity and
  gradient shear bands in wormlike micellar solutions of different cpcl/salt
  systems, Journal of Rheology 58~(6) (2014) 1647--1672.

\bibitem{Dhont2008}
J.~K. Dhont, W.~J. Briels, Gradient and vorticity banding, Rheologica acta
  47~(3) (2008) 257--281.

\bibitem{Olmsted1999}
P.~D. Olmsted, Two-state shear diagrams for complex fluids in shear flow, EPL
  (Europhysics Letters) 48~(3) (1999) 339.

\bibitem{Decruppe1995}
J.~Decruppe, R.~Cressely, R.~Makhloufi, E.~Cappelaere, Flow birefringence
  experiments showing a shear-banding structure in a ctab solution, Colloid and
  Polymer Science 273~(4) (1995) 346--351.

\bibitem{Olmsted2008}
P.~D. Olmsted, Perspectives on shear banding in complex fluids, Rheologica Acta
  47~(3) (2008) 283--300.

\bibitem{Yerushalmi1970}
J.~Yerushalmi, S.~Katz, R.~Shinnar, The stability of steady shear flows of some
  viscoelastic fluids, Chemical Engineering Science 25~(12) (1970) 1891--1902.

\bibitem{Cromer2013}
M.~Cromer, M.~C. Villet, G.~H. Fredrickson, L.~G. Leal, Shear banding in
  polymer solutions, Physics of Fluids 25~(5) (2013) 051703.

\bibitem{Divoux2016}
T.~Divoux, M.~A. Fardin, S.~Manneville, S.~Lerouge, Shear banding of complex
  fluids, Annual Review of Fluid Mechanics 48 (2016) 81--103.

\bibitem{Petrie1976}
C.~J. Petrie, M.~M. Denn, Instabilities in polymer processing, AIChE Journal
  22~(2) (1976) 209--236.

\bibitem{Olmsted1999phase}
P.~Olmsted, C.~David~Lu, Phase coexistence of complex fluids in shear flow,
  Faraday Discussions 112 (1999) 183--194.

\bibitem{Fielding2007}
S.~M. Fielding, Vorticity structuring and velocity rolls triggered by gradient
  shear bands, Physical Review E 76~(1) (2007) 016311.

\bibitem{Chacko2018}
R.~N. Chacko, R.~Mari, M.~E. Cates, S.~M. Fielding, Dynamic vorticity banding
  in discontinuously shear thickening suspensions, Physical Review Letters
  121~(10) (2018) 108003.

\bibitem{Boltenhagen1997_2}
P.~Boltenhagen, Y.~Hu, E.~Matthys, D.~Pine, Inhomogeneous structure formation
  and shear-thickening in worm-like micellar solutions, EPL (Europhysics
  Letters) 38~(5) (1997) 389.

\bibitem{Liu1996}
C.-H. Liu, D.~Pine, Shear-induced gelation and fracture in micellar solutions,
  Physical Review Letters 77~(10) (1996) 2121.

\bibitem{Hu1998}
Y.~Hu, P.~Boltenhagen, D.~Pine, Shear thickening in low-concentration solutions
  of wormlike micelles. i. direct visualization of transient behavior and phase
  transitions, Journal of Rheology 42~(5) (1998) 1185--1208.

\bibitem{Wilson1992}
G.~M. Wilson, B.~Khomami, An experimental investigation of interfacial
  instabilities in multilayer flow of viscoelastic fluids: Part 1. incompatible
  polymer systems, Journal of Non-Newtonian Fluid Mechanics 45~(3) (1992)
  355--384.

\bibitem{Wilson1993_1}
G.~M. Wilson, B.~Khomami, An experimental investigation of interfacial
  instabilities in multilayer flow of viscoelastic fluids. part 2. elastic and
  nonlinear effects in incompatible polymer systems, Journal of Rheology 37~(2)
  (1993) 315--339.

\bibitem{Wilson1993_2}
G.~M. Wilson, B.~Khomami, An experimental investigation of interfacial
  instabilities in multilayer flow of viscoelastic fluids. 3. compatible
  polymer systems, Journal of Rheology 37~(2) (1993) 341--354.

\bibitem{Yamani2023}
S.~Yamani, Y.~Raj, T.~A. Zaki, G.~H. McKinley, I.~Bischofberger, Spatiotemporal
  signatures of elastoinertial turbulence in viscoelastic planar jets, Physical
  Review Fluids 8~(6) (2023) 064610.

\bibitem{Turner1992}
M.~Turner, M.~Cates, Flow-induced phase transitions in rod-like micelles,
  Journal of Physics: Condensed Matter 4~(14) (1992) 3719.

\bibitem{CatesTurner1990}
M.~Cates, M.~Turner, Flow-induced gelation of rodlike micelles, EPL
  (Europhysics Letters) 11~(7) (1990) 681.

\bibitem{Vasquez2007}
P.~A. Vasquez, G.~H. McKinley, L.~P. Cook, A network scission model for
  wormlike micellar solutions: I. model formulation and viscometric flow
  predictions, Journal of Non-Newtonian Fluid Mechanics 144~(2-3) (2007)
  122--139.

\bibitem{Dutta2018}
S.~Dutta, M.~D. Graham, {Mechanistic constitutive model for wormlike micelle
  solutions with flow-induced structure formation}, Journal of Non-Newtonian
  Fluid Mechanics 251 (2018) 97--106.
\newblock \href {http://arxiv.org/abs/1711.08358} {\path{arXiv:1711.08358}},
  \href {https://doi.org/10.1016/j.jnnfm.2017.12.001}
  {\path{doi:10.1016/j.jnnfm.2017.12.001}}.

\bibitem{Bautista1999}
F.~Bautista, J.~De~Santos, J.~Puig, O.~Manero, Understanding thixotropic and
  antithixotropic behavior of viscoelastic micellar solutions and liquid
  crystalline dispersions. 1. the model, Journal of Non-Newtonian Fluid
  Mechanics 80~(2-3) (1999) 93--113.

\bibitem{Fredrickson1970}
A.~Fredrickson, A model for the thixotropy of suspensions, AIChE Journal 16~(3)
  (1970) 436--441.

\bibitem{Manero2007}
O.~Manero, J.~P{\'e}rez-L{\'o}pez, J.~Escalante, J.~Puig, F.~Bautista, A
  thermodynamic approach to rheology of complex fluids: The generalized bmp
  model, Journal of non-newtonian fluid mechanics 146~(1-3) (2007) 22--29.

\bibitem{Lopez2018}
J.~L{\'o}pez-Aguilar, M.~Webster, H.~Tamaddon-Jahromi, O.~Manero, Predictions
  for circular contraction-expansion flows with viscoelastoplastic \&
  thixotropic fluids, Journal of Non-Newtonian Fluid Mechanics 261 (2018)
  188--210.

\bibitem{Lopez2022}
J.~E. Lopez-Aguilar, O.~Resendiz-Tolentino, H.~R. Tamaddon-Jahromi, M.~Ellero,
  O.~Manero, Flow past a sphere: Numerical predictions of
  thixo-viscoelastoplastic wormlike micellar solutions, Journal of
  Non-Newtonian Fluid Mechanics 309 (2022) 104902.

\bibitem{Tamano2020}
S.~Tamano, S.~Hamanaka, Y.~Nakano, Y.~Morinishi, T.~Yamada, Rheological
  modeling of both shear-thickening and thinning behaviors through constitutive
  equations, Journal of Non-Newtonian Fluid Mechanics 283 (2020) 104339.

\bibitem{Shekar2020}
A.~Shekar, R.~M. McMullen, B.~J. McKeon, M.~D. Graham, Self-sustained
  elastoinertial tollmien--schlichting waves, Journal of Fluid Mechanics 897
  (2020) A3.

\bibitem{Dubief2023}
Y.~Dubief, V.~E. Terrapon, B.~Hof, Elasto-inertial turbulence, Annual Review of
  Fluid Mechanics 55 (2023).

\bibitem{Larson1999}
R.~G. Larson, {Structure and Rheology of Complex Fluids}, Oxford University
  Press, 1999.

\bibitem{Doi1986}
M.~Doi, S.~F. Edwards, {The theory of polymer dynamics}, Claredon Press, Oxford
  (1986).
\newblock \href {https://doi.org/10.1016/S1359-0286(96)80106-9}
  {\path{doi:10.1016/S1359-0286(96)80106-9}}.

\bibitem{Dhont2006}
J.~K. Dhont, W.~J. Briels, Rod-like brownian particles in shear flow, Soft
  Matter: Complex Colloidal Suspensions, edited by G. Gompper, M. Schick 2
  (2006).

\bibitem{Forest2003}
M.~G. Forest, Q.~Wang, Monodomain response of finite-aspect-ratio
  macromolecules in shear and related linear flows, Rheologica Acta 42~(1)
  (2003) 20--46.

\bibitem{Dhont2003}
J.~K. Dhont, W.~J. Briels, {Viscoelasticity of suspensions of long, rigid
  rods}, Colloids and Surfaces A: Physicochemical and Engineering Aspects
  213~(2-3) (2003) 131--156.
\newblock \href {https://doi.org/10.1016/S0927-7757(02)00508-3}
  {\path{doi:10.1016/S0927-7757(02)00508-3}}.

\bibitem{Graham:2018ty}
M.~D. Graham, {Microhydrodynamics, Brownian Motion, and Complex Fluids},
  Cambridge Texts in Applied Mathematics, Cambridge University Press,
  Cambridge, 2018.

\bibitem{Jasak2007}
H.~Jasak, A.~Jemcov, Z.~Tukovic, et~al., Openfoam: A c++ library for complex
  physics simulations, in: International workshop on coupled methods in
  numerical dynamics, Vol. 1000, 2007, pp. 1--20.

\bibitem{Pimenta2017}
F.~Pimenta, M.~Alves, Stabilization of an open-source finite-volume solver for
  viscoelastic fluid flows, Journal of Non-Newtonian Fluid Mechanics 239 (2017)
  85--104.

\bibitem{rheoTool}
F.~Pimenta, M.~Alves, rheotool, \url{https://github.com/fppimenta/rheoTool}
  (2016).

\bibitem{Favero2010}
J.~Favero, A.~Secchi, N.~Cardozo, H.~Jasak, Viscoelastic flow analysis using
  the software openfoam and differential constitutive equations, Journal of
  Non-Newtonian Fluid Mechanics 165~(23-24) (2010) 1625--1636.

\bibitem{Lu2000}
C.-Y.~D. Lu, P.~D. Olmsted, R.~Ball, Effects of nonlocal stress on the
  determination of shear banding flow, Physical Review Letters 84~(4) (2000)
  642.

\end{thebibliography}

\section{Appendix A: Supplemental material}
Movie S1. Micelle length in start-up of steady shear flow with $\mathrm{Pe} = 0.0225$ in 2D.

Movie S2. Micelle length for a decrease in shear rate from $\mathrm{Pe} = 0.04$ to  $\mathrm{Pe} = 0.015$ in 2D.

Movie S3. Micelle length in start-up of steady shear flow with $\mathrm{Pe} = 0.0225$ in 3D. Left: $z = 0$, right: $z = \pm h/2$, bottom: $rz$-plane at $\theta = \pi$.

Movie S4. Micelle surfaces in start-up of steady shear flow with $\mathrm{Pe} = 0.0225$ in 3D. Micelle structures shown for $L \geq 5$.

Movie S5. Micelle length for a decrease in shear rate from $\mathrm{Pe} = 0.04$ to  $\mathrm{Pe} = 0.015$ in 3D. Left: $z = 0$, right: $z = \pm h/2$, bottom: $rz$-plane at $\theta = \pi$.

\end{document}